\documentclass[11pt]{article}

\usepackage{amssymb,amsmath,amsthm}
\usepackage{algorithm,algorithmicx,algpseudocode}
\usepackage{tikz, verbatim, color, tabularx, booktabs}
\usepackage[margin=1in]{geometry}
\usepackage[margin=.1in,font=small]{caption}
\usetikzlibrary{shapes,arrows,positioning}
\usepackage[outdir=./]{epstopdf}
\usepackage[strict]{changepage}

\algnewcommand\algorithmicswitch{\textbf{switch}}
\algnewcommand\algorithmiccase{\textbf{case}}
\algdef{SE}[SWITCH]{Switch}{EndSwitch}[1]{\algorithmicswitch\ #1}{\algorithmicend\ \algorithmicswitch}
\algdef{SE}[CASE]{Case}{EndCase}[1]{\algorithmiccase\ #1}{\algorithmicend\ \algorithmiccase}
\algtext*{EndCase}

\newtheorem{theorem}{Theorem}
\newtheorem{definition}[theorem]{Definition}

\newtheorem{remark}{Remark}
\newtheorem{proposition}[theorem]{Proposition}
\newtheorem{corollary}[theorem]{Corollary}

\newcommand{\reals}{\mathbb{R}}											
\newcommand{\sgn}{\text{sgn}}											
\newcommand{\prob}{\mathbb{P}}											
\newcommand{\expe}{\mathbb{E}}											
\newcommand{\todo}[1]{}													

\newcommand{\longonly}[1]{{#1}}											
\newcommand{\shortonly}[1]{{#1}}										

\renewcommand{\shortonly}[1]{}											

\title{Gradual Release of Sensitive Data under Differential Privacy}
\author{Fragkiskos Koufogiannis, Shuo Han, George J. Pappas\footnote{Department of Electrical and Systems Engineering, University of Pennsylvania. Emails:  \texttt{\{fkouf,hanshuo,pappasg\}@seas.upenn.edu}. This work was supported in part by the TerraSwarm Research Center, one of six centers supported by the STARnet phase of the Focus Center Research Program (FCRP) a Semiconductor Research Corporation program sponsored by MARCO and DARPA.} }

\begin{document}

\maketitle
\thispagestyle{empty}

\begin{abstract}
We introduce the problem of releasing sensitive data under differential privacy when the privacy level is subject to change over time. Existing work assumes that privacy level is determined by the system designer as a fixed value before sensitive data is released. For certain applications, however, users may wish to relax the privacy level for subsequent releases of the same data after either a re-evaluation of the privacy concerns or the need for better accuracy. Specifically, given a database containing sensitive data, we assume that a response $y_1$ that preserves $\epsilon_{1}$-differential privacy has already been published. Then, the privacy level is relaxed to $\epsilon_2$, with $\epsilon_2 > \epsilon_1$, and we wish to publish a more accurate response $y_2$ while the joint response $(y_1, y_2)$ preserves $\epsilon_2$-differential privacy. How much accuracy is lost in the scenario of gradually releasing two responses $y_1$ and $y_2$ compared to the scenario of releasing a single response that is $\epsilon_{2}$-differentially private? Our results show that there exists a composite mechanism that achieves \textit{no loss} in accuracy. 
  
We consider the case in which the private data lies within $\reals^{n}$ with an adjacency relation induced by the $\ell_{1}$-norm, and we focus on mechanisms that approximate identity queries. We show that the same accuracy can be achieved in the case of gradual release through a mechanism whose outputs can be described by a \textit{lazy Markov stochastic process}. This stochastic process has a closed form expression and can be efficiently sampled. Our results are applicable beyond identity queries. To this end, we demonstrate that our results can be applied in several cases, including Google's RAPPOR project, trading of sensitive data, and controlled transmission of private data in a social network. Finally, we conjecture that gradual release of data \textit{without performance loss} is an intrinsic property of differential privacy and, thus, holds in more general settings.

\end{abstract}

\clearpage
\setcounter{page}{1}

\section{Introduction}
Differential privacy is a framework that provides rigorous privacy guarantees for the release of sensitive data. The intrinsic trade-off between the privacy guarantees and accuracy of the privacy-preserving mechanism is controlled by the privacy level $\epsilon \in [0,\infty)$; smaller values of $\epsilon$ imply stronger privacy and less accuracy. Specifically, \textit{end users}, who are interested in the output of the mechanism, demand acceptable accuracy of the privacy-preserving mechanism, whereas, \textit{owners} of sensitive data are interested in strong enough privacy guarantees.

Existing work on differential privacy assumes that the privacy level is determined prior to release of any data and remains constant throughout the life of the privacy-preserving mechanism. However, for certain applications, the privacy level may need to be revised \textit{after} data has been released, due to either users' need for improved accuracy or after owners' re-evaluation of the privacy concerns.
One such application is trading of private data, where the owners re-evaluate their privacy concerns after monetary payments. Specifically, the end users initially access private data under $\epsilon_{1}$ privacy guarantees and they later decide to \textit{``buy''} more accurate data, relax privacy level to $\epsilon_{2}$, and enjoy better accuracy.
Furthermore, the need for more accurate responses may dictate a change in the privacy level. In particular, a database containing sensitive data is persistent over time; e.g. a database of health records contains the same patients with the same health history over several years. Future uses of the database may require better accuracy, especially, after a threat is suspected (e.g. virus spread, security breach). These two example applications share the same core questions.

Is it possible to release a preliminary response with $\epsilon_{1}$-privacy guarantees and, later, release a more accurate and less private response with overall $\epsilon_{2}$-privacy guarantees? How is this scenario compared to publishing a single response under $\epsilon_{2}$-privacy guarantees? In fact, is the performance of the second response damaged by the preliminary one?
 
Composition theorems \cite{mcsherry07} provide a simple, but suboptimal, solution to gradually releasing sensitive data. Given an initial privacy level $\epsilon_{1}$, a noisy, privacy-preserving response $y_{1}$ is generated. Later, the privacy level is increased to a new value $\epsilon_{2}$ and a new response $y_{2}$ is published. For an overall privacy level of $\epsilon_{2}$, the second response $y_{2}$ needs to be $(\epsilon_{2}-\epsilon_{1})$-private, according to the composition theorem. Therefore, the accuracy of the second response deteriorates because of the initial release $y_{1}$.

In this work, we derive a composite mechanism which exhibits \textit{no loss} in accuracy after the privacy level is relaxed. This mechanism employs correlation between successive responses, and, to the best of our knowledge, is the first mechanism that performs gradual release of sensitive data.

\subsection{Our Results}
This work introduces the problem of gradually releasing sensitive data. Our results focus on the case of vector-valued sensitive data $u\in\reals^{n}$ with an $\ell_{1}$-norm adjacency relation. Our first result states that, for the one-dimensional ($n=1$) identity query, there is an algorithm which relaxes privacy in two steps without sacrificing any accuracy. Although our technical treatment focuses on identical queries, our results are applicable to a broader family of queries. We also prove the \textit{Markov property} for this algorithm and, thus, we can easily (without any computational complexity) relax privacy in any number of steps. These two results provide a different perspective of differential privacy, and lead to the definition of a \textit{lazy} Markov stochastic process indexed by the privacy level $\epsilon$. Gradually releasing sensitive data is performed by sampling once from this stochastic process. We also extend the results to the high-dimensional case.

On a theoretical level, our contributions add a whole new dimension to differential privacy --- that of a \textit{varying} parameter $\epsilon$. We focus on the mechanism that adds Laplace-distributed noise $V_{\epsilon}$ to the private data $u\in\reals^{n}$:
\begin{align} \label{eqn:laplaceMechanism:intro} 
	Q_{\epsilon}u = u + \begin{bmatrix} v_{\epsilon}^{(1)} \\ \vdots \\ v_{\epsilon}^{(n)} \end{bmatrix},
\end{align}
where $\epsilon$ is the privacy level, $\|\cdot\|_{1}$ is the $\ell_{1}$-norm, and $\{ v_{\epsilon}^{(i)} \}_{i=1}^{n}$ are independent and identically distributed samples from the stochastic process $\{ V_{\epsilon} \}_{\epsilon>0}$ which has the following properties:
\begin{enumerate}
	\item $\{ V_{\epsilon} \}_{\epsilon>0}$ is Markov: $ V_{\epsilon_{1}} \bot V_{\epsilon_{3}}  | V_{\epsilon_{2}}, \text{ for any } \epsilon_{3} \geq \epsilon_{2} \geq \epsilon_{1} > 0 $.
	\item $V_{\epsilon}$ is Laplace-distributed: $ \prob( V_{\epsilon} = x )  = \frac{\epsilon}{2} e^{-\epsilon |x|} $.
	\item $\{ V_{\epsilon} \}_{\epsilon>0}$ is \textit{lazy}, i.e. there is positive probability of not changing value):
		$$ \prob( V_{\epsilon_{1}} = x | V_{\epsilon_{2}} = y ) =  \left( \frac{\epsilon_{1}}{\epsilon_{2}} \right)^{2}  \delta(x-y) + \left( 1 - \left( \frac{\epsilon_{1}}{\epsilon_{2}} \right)^{2} \right) \frac{ \epsilon_{1} }{2} e^{-\epsilon_{1} |x-y|}, \text{ where } \epsilon_{2} \geq \epsilon_{1} > 0, $$
	where $\delta$ is Dirac's delta function. For a fixed $\epsilon$, mechanism~\eqref{eqn:laplaceMechanism:intro} reduces to the Laplace mechanism.
\end{enumerate}

Mechanism \eqref{eqn:laplaceMechanism:intro} has the following properties and, thus, performs gradual release of private data:
\begin{itemize}
	\item \textbf{Privacy:} For any set of privacy levels $\{ \epsilon_{i} \}_{i=1}^{m}$, the mechanism that responds with $\{ Q_{\epsilon_{i}}u \}_{i=1}^{m} $ is $\left( \max_{i=1}^{m} \epsilon_{i} \right)$-private.
	\item \textbf{Accuracy:} For a fixed $\epsilon$, the mechanism $Q_{\epsilon}$ is the optimal $\epsilon$-private mechanism.
\end{itemize}
In practice, gradual release of private data is achieved by sampling the stochastic process $\{ V_{\epsilon} \}_{\epsilon>0}$:
\begin{enumerate}
	\item Draw a single sample $\{ v_{\epsilon} \}_{\epsilon>0}$ from the stochastic process $\{ V_{\epsilon} \}_{\epsilon>0}$.
	\item Compute the signal $y_{\epsilon} = u + v_{\epsilon}$, $\epsilon>0$.
	\item For $\epsilon_{1}$-privacy guarantees, release the random variable $y_{\epsilon_{1}}$.
	\item Once privacy level is relaxed from $\epsilon_{1}$ to $\epsilon_{2}$, where $\epsilon_{2} \geq \epsilon_{1}$, release the random variable $y_{\epsilon_{2}}$.
	\item In order to relax privacy level in an arbitrarily many times, $\epsilon_{1} \to \epsilon_{2} \to \cdots \to \epsilon_{m}$, repeat the last step.
\end{enumerate}

More formally, our main result derives a composite mechanism that gradually releases private data by relaxing the privacy level in an arbitrary number of steps.
\begin{theorem}[\longonly{\textbf{A.} }Gradual Privacy as a Composite Mechanism] \label{thm:mainResultInformal}
	Let $\reals^{n}$ be the space of privacy data equipped with an $\ell_{1}$-norm adjacency relation. Consider $m$ privacy levels $\{\epsilon_{i}\}_{i=1}^{m}$ such that $0\leq\epsilon_{1}\leq\cdots\leq\epsilon_{m}$ which successively relax the privacy level. Then, there exists a composite mechanism $Q$ of the form
	\begin{align} \label{eqn:compositeMechanism:intro}
		Qu := \left(u+V_{1}, \ldots, u+V_{m} \right),
	\end{align}
	such that:
	\begin{enumerate}
		\item The restriction of the mechanism $Q$ to the first $j$ coordinates $(u+V_{1}, \ldots, u+V_{j})$ is $\epsilon_{j}$-private, for any $j\in\{1,\ldots,m\}$.
		\item Each coordinate $j\in\{1,\ldots,m\}$ of the mechanism $u+V_{j}$ achieves the optimal mean-squared error $\expe \| V_{j} \|_{2}^{2} $.
	\end{enumerate}
\end{theorem}

\longonly{
The mechanism that satisfies Theorem \ref{thm:mainResultInformal} has a closed-form expression and provides a new perspective of differential privacy. Instead of designing composite mechanisms of the form \eqref{eqn:compositeMechanism:intro}, we consider the \textit{continuum} of privacy levels $\epsilon\in[0,\infty)$. Our results are more succinctly stated in terms of a stochastic process $\{ V_{\epsilon} \}_{\epsilon>0}$. A composite mechanism is recovered from the stochastic process by sampling the process $\{ V_{\epsilon} \}_{\epsilon>0}$ at a finite set of privacy levels $\{ \epsilon_{i} \}_{i=1}^{m}$.
\addtocounter{theorem}{-1}
\begin{theorem}[\textbf{B.} Gradual Privacy as a Stochastic Process] \label{thm:mainResultProcess}
	Let $\reals^{n}$ be the space of privacy data equipped with the $\ell_{1}$-norm. Then, there exists a stochastic process $\{ V_{\epsilon} \}_{\epsilon>0} \}$ that defines the family of mechanisms $Q_{\epsilon}$ parametrized by $\epsilon$:
	\begin{align}
		Q_{\epsilon}u := u + V_{\epsilon}, \quad \epsilon \in (0,\infty),
	\end{align}
	such that:
	\begin{itemize}
		\item \textbf{Privacy:} For any $\epsilon>0$, the mechanism that releases the signal $\{ u+V_{\sigma} \}_{\sigma\in(0,\epsilon]}$ is $\epsilon$-private.
		\item \textbf{Accuracy:} The mechanism $Q_{\epsilon}$ that releases the random variable $u+ V_{\epsilon}$ is the optimal $\epsilon$-private mechanism, i.e. the noise sample $V_{\epsilon}$  achieves the optimal mean-squared error  $\expe \| V_{\epsilon} \|_{2}^{2}$.
	\end{itemize}
\end{theorem}
}

\begin{figure} \begin{center}
\includegraphics[width=\linewidth]{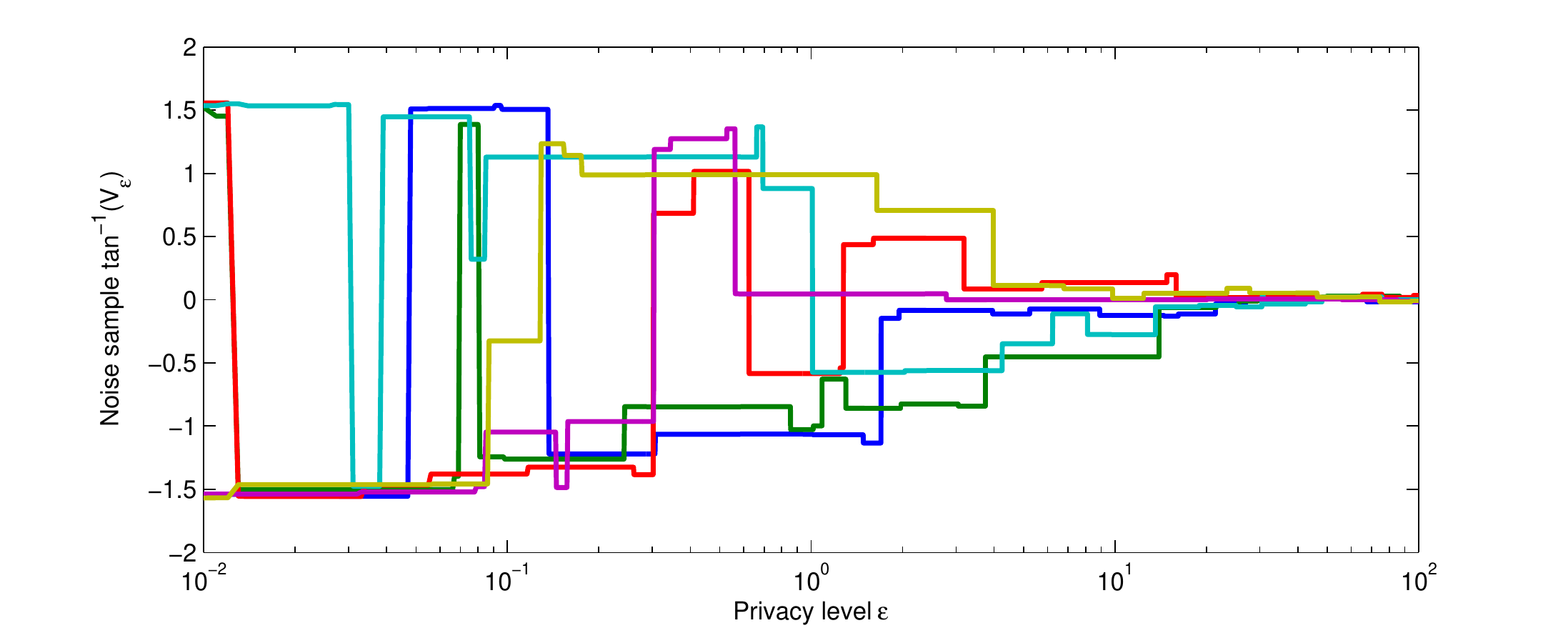}
\caption{Gradual release of identity queries is achieved with the use of the stochastic process $V_{\epsilon}$ for $\epsilon\geq0$. For tight values of privacy ($\epsilon\to0$), high values of noise ($\left| \tan^{-1}V_{\epsilon} \right|\to\frac{\pi}{2}$) are returned, whereas, almost zero samples ($V_{\epsilon}\to0$) are returned for large privacy budgets ($\epsilon\to\infty$). The process $V_{\epsilon}$ is Markov; future samples depend only on the current value of the process which eases implementation. Furthermore, the process is lazy; the value of the process changes only a few times.}  \label{fig:privateProcess1D}
\end{center} \end{figure}

From a more practical point of view, our results are applicable to cases beyond identity queries. Specifically, our results are directly applicable to a broad family of privacy-preserving mechanisms that are built upon the Laplace mechanism and, informally, have the following form. The sensitive data is initially preprocessed, then, the Laplace mechanism is invoked, and, finally, a post-processing step occurs. Under the assumption that the preprocessing step is invariant of the privacy level, gradual release of sensitive data is possible. We demonstrate the applicability of our results on Google's RAPPOR project \cite{erlingsson2014rappor}, which analyzes software features that individuals use while respecting their privacy. In particular, if a software feature is suspected to be malicious, privacy level can be gradually relaxed and a more accurate analysis can be performed. On another direction, our results broaden the spectrum of applications of differential privacy. To this end, we present an application to social networks where users have different privacy concerns against close friends, acquaintances, and strangers.

We conclude our paper with a conjecture. Although present work focuses on mechanisms that add Laplace-distributed noise, we conjecture that the feasibility of gradually releasing sensitive data is a more general property of differential privacy. In particular, we formulate the conjecture that repeatedly relaxing the privacy level without loss of accuracy is possible for a larger family of privacy-aware mechanisms.

\subsection{Previous Work}
Differential privacy is an active field of research and a rich spectrum of differential private mechanisms has appeared in the literature. The exponential mechanism \cite{mcsherry07} is a powerful and generic tool for building differential private mechanisms. In particular, mechanisms that efficiently approximate linear (counting) queries have received a lot of attention \cite{hardt10}, \cite{li10}, \cite{ullman2013answering}. Besides counting queries, privacy-aware versions of more complex quantities have been introduced such as signal filtering \cite{le2014differentially}, optimization problems \cite{gupta2010differentially}, \cite{han2014differentially}, and allocation problems \cite{hsu2014private}. In addition to the theoretical work, differential privacy has been deployed in software tools \cite{reed10}.

The aforementioned work assumes that the privacy level $\epsilon$ is a designer's choice that is held fixed throughout the life of the privacy-aware mechanism. To the best of our knowledge, our work is the first approach that considers privacy-aware mechanisms with a varying privacy level $\epsilon$. Gradually releasing private data resembles the setting of differential privacy under continuous observation, which was first studied in \cite{dwork10}. In that setting of \cite{dwork10}, the privacy level remains fixed while more sensitive data is being added to the database and more responses are released. In contrast, our setting assumes that both the sensitive data and the quantity of interest are fixed and the privacy level $\epsilon$ is varying.

Gradual release of sensitive data is closely related to optimality results. Work in \cite{hardt10} established optimality results in an asymptotic sense (with the size of the database). Instead, our work requires \textit{exact} optimality results and, therefore, is presented within a tighter version of differential privacy that was explored in \cite{chatzikokolakis13}, \cite{koufogiannis15laplaceOptimality}, where exact optimality results exist. This tighter notion which is targeted for metric spaces and we call \textit{Lipschitz privacy}, allows for the use of optimization techniques and calculus tools. Prior work on Lipschitz privacy includes the \textit{exact} optimality of the Laplace mechanism is established under Lipschitz privacy \cite{koufogiannis15laplaceOptimality}, \cite{wang14}.

On a more technical level, most prior work on differential privacy \cite{gupta2010differentially}, \cite{han2014differentially}, \cite{hsu2014private} introduces differential private mechanisms that are built upon the Laplace mechanism and variations of it. Although building upon the Laplace mechanism limits the solution space, there is a good reason for doing so. Specifically, for non-trivial applications, the space of probability measures can be extremely rich and hard to deal with. Technically, our approach deviates from prior work by searching over the whole space of differential private mechanisms. Work in \cite{xiao2011ireduct} is another example that proposes a non-Laplace distribution in order to achieve better performance on subsequent queries while satisfying overall differential privacy constraints. The Laplace mechanism, then, naturally emerges as the optimal mechanism.

\section{Background Information}
\subsection{Differential Privacy}
\longonly{
The framework of differential privacy \cite{dwork06}, \cite{dwork2013algorithmic} dictates that, whenever sensitive data is accessed, a noisy response is returned. The statistics of the injected noise are deliberately designed to ensure two things. First, an adversary that observes the noisy response cannot \textit{confidently infer} the original sensitive data. The privacy level is parametrized by $\epsilon\in[0,\infty)$, where smaller values of $\epsilon$ imply stronger privacy guarantees. Second, the noisy response can still be used as a surrogate of the exact response \textit{without severe performance degradation}. On the other hand, the accuracy of the noisy response is quantified by the mean-squared error from the exact response.
}

Work in \cite{dwork06}\shortonly{, \cite{dwork2013algorithmic}} defined differential privacy, which provides strong privacy guarantees against a powerful adversary.
\begin{definition}[Differential Privacy]
Let $\mathcal{U}$ be a set of private data, $\mathcal{A}\subseteq\mathcal{U}^{2}$ be a symmetric binary relation (called adjacency relation) and $\mathcal{Y}$ be a set of possible responses. For $\epsilon>0$, the randomized mapping $Q: \mathcal{U} \rightarrow \Delta\left( \mathcal{Y} \right)$ (called mechanism) is $\epsilon$-differentially private if
\begin{align}
\prob( Qu\in\mathcal{S} ) \leq e^{\epsilon} \prob(Qu'\in\mathcal{S} ), \; \forall (u,u')\in\mathcal{A}, \; \forall \mathcal{S}\subseteq\mathcal{Y}.
\end{align}
\end{definition}

\longonly{
\begin{remark}
We assume the existence of a rich-enough $\sigma$-algebra $M\subseteq2^{\mathcal{Y}}$ on the set of possible responses $\mathcal{Y}$. Then, $\Delta\left( \mathcal{Y} \right)$ denotes the set of probability measures over $(M,\mathcal{Y})$.
\end{remark}
}

Let $y\sim Qu$ be a noisy response produced by the $\epsilon$-differentially private mechanism $Q$. For brevity, we say that ``output $y$ preserves $\epsilon$-privacy of the input $u$''.

The adjacency relation $\mathcal{A}$ captures the aspects of the private data $u$ that are deemed sensitive. Consider a scheme with $n$ users, where each user $i$ contributes her real-valued private data $u_{i}\in\reals$, and a private database $u=[u_{1},\dots,u_{n}]\in\reals^{n}$ is composed. For $\alpha>0$, an adjacency relation that captures the participation of a single individual to the aggregating scheme is defined as:
\begin{align} \label{eqn:L0Adjacency}
	(u,u') \in \mathcal{A}_{\ell_{0}} \Leftrightarrow \exists j \text{ s.t. } u_{i}=u_{i}', \forall i\neq j \text{ and } |u_{j}-u_{j}'|\leq \alpha.
\end{align}
Adjacency relation $\mathcal{A}_{\ell_{0}}$ can be relaxed to $\mathcal{A}_{\ell_{1}}$, which is induced by the $\ell_{1}$-norm and is defined as:
\begin{align} \label{eqn:L1Adjacency}
	(u,u') \in \mathcal{A}_{\ell_{1}} \Leftrightarrow \|u - u\|_{1} \leq \alpha,
\end{align}
where it holds that $\mathcal{A}_{\ell_{0}} \subseteq \mathcal{A}_{\ell_{1}}$.

\longonly{
Resilience to post-processing establishes that any post-processing on the output of an $\epsilon$-differentially private mechanism cannot hurt the privacy guarantees.
\begin{proposition}[Resilience to Post-Processing] \label{thm:postprocessing}
Let $Q:\mathcal{U} \to \Delta\left( \mathcal{Y} \right)$ be an $\epsilon$-differentially private mechanism and $g:\mathcal{Y} \to \mathcal{Z}$ be a possibly randomized function. Then, the mechanism $g \circ Q$ is also $\epsilon$-differentially private.
\end{proposition}
}

More complicated mechanisms can be defined from simple ones using the composition theorem.
\begin{proposition}[Composition] \label{thm:composition}
Let mechanisms $Q_{1}, Q_{2}: \mathcal{U} \to \Delta \left( \mathcal{Y} \right)$ respectively satisfy $\epsilon_{1}$ and $\epsilon_{2}$-differential privacy. Then, the composite mechanism $Q: \mathcal{U} \to \Delta \left( \mathcal{Y}^{2} \right)$ defined by $Q = (Q_{1},Q_{2})$ is $(\epsilon_{1}+\epsilon_{2})$-differentially private.
\end{proposition}

Proposition \ref{thm:composition} provides privacy guarantees whenever the \textit{same} sensitive data is repeatedly used. Moreover, the resulting privacy level $\epsilon_{1}+\epsilon_{2}$ given by Proposition \ref{thm:composition} is an upper bound and can severely over-estimate the actual privacy level. The mechanism presented in this paper introduces correlation between mechanisms $Q_{1}$ and $Q_{2}$, so that it provides much stronger privacy guarantees.

\subsection{Lipschitz Privacy}
Lipschitz privacy \cite{chatzikokolakis13}, \cite{koufogiannis15laplaceOptimality} is a slightly stronger version of differential privacy and is often used when the data is defined on metric spaces. 

\begin{definition}[Lipschitz Privacy] \label{def:lipschitzPrivacy}
Let $\left( \mathcal{U}, d\right)$ be a metric space and $\mathcal{Y}$ be the set of possible responses. For $\epsilon>0$, the mechanism $Q$ is $\epsilon$-Lipschitz private if the following Lipschitz condition holds:
	\begin{align} \label{eqn:lipschitzConstraint}
		\left| \ln \prob( Qu\in\mathcal{S} ) - \ln \prob( Qu\in\mathcal{S} ) \right| \leq \epsilon d(u,u'), \quad \forall u,u'\in\mathcal{U}, \ \forall \mathcal{S}\subseteq\mathcal{Y}.
	\end{align}
\end{definition}

Lipschitz privacy is closely related to the original definition of differential privacy, where the adjacency relation $\mathcal{A}$ in differential privacy is defined through the metric $d$. In fact, any Lipschitz private mechanism is also differentially private.
\begin{proposition} \label{thm:lipschitzDifferential}
For any $\alpha>0$, an $\epsilon$-Lipschitz private mechanism $Q$ is $\alpha \epsilon$-differentially private under the adjacency relation $\mathcal{A}$:
	\begin{align}
		(u,u') \in \mathcal{A} \Leftrightarrow  d(u,u')\leq \alpha.
	\end{align}
\end{proposition}

Adjacency relation $\mathcal{A}_{\ell_{1}}$ defined in \eqref{eqn:L1Adjacency} can be captured by the $\ell_{1}$-norm under the notion of Lipschitz privacy; the metric $d$ is $d(u,u') = \|u - u' \|_{1}$.

Our results are stated within the Lipschitz privacy framework. Proposition \ref{thm:lipschitzDifferential} implies that our privacy results remain valid within the framework of differential privacy. For brevity, we call an $\epsilon$-Lipschitz private mechanism as $\epsilon$-private and imply that a differentially private mechanism can be derived.

Similar to differential privacy, Lipschitz privacy is preserved under post-processing (Proposition~\ref{thm:postprocessing}) and composition of mechanisms (Proposition~\ref{thm:composition}). Compared to differential privacy, Lipschitz privacy is more convenient to work with when the data and adjacency relation are defined on a metric space, which allows for the use of calculus tools. Under mild assumptions, the Lipschitz constraint \eqref{eqn:lipschitzConstraint} is equivalent to a derivative bound. In particular, for $\mathcal{U} = \reals^{n}$ equipped with the metric induced by the norm $\|\cdot\|$, a mechanism $Q$ is $\epsilon$-Lipschitz private if
\begin{align} \label{eqn:derivativeBound}
	\left\| \nabla_{u} \ln \prob( Qu = y ) \right\|_{*} \leq \epsilon,
\end{align}
where $\|\cdot\|_{*}$ is the dual norm of $\|\cdot\|$. In practice, we check condition \eqref{eqn:derivativeBound} to establish the privacy properties of mechanism $Q$.

\subsection{Optimality of the Laplace Mechanism}
Computing the optimal private mechanism for a fixed privacy level $\epsilon$ is considered an open problem. \longonly{Specifically, let $\mathcal{U}$ be the space of private data, $\mathcal{A}$ be an adjacency relation, $q:\mathcal{U}\to\mathcal{Y}$ be a query, and $\epsilon$ be a fixed privacy level.
The exponential mechanism \cite{mcsherry07} is a popular technique for constructing private mechanisms. 
\begin{proposition}[Exponential Mechanism]
Let $s:\mathcal{U} \times \mathcal{Y} \to \reals$ be $1$-Lipschitz in $(\mathcal{U},d)$. Consider the mechanism $Q$ whose output satisfies
	\begin{align}
		\prob( Qu = y ) \propto e^{ \epsilon s(u,y) }.
	\end{align}
Then, $Q$ is $\epsilon$-Lipschitz private.
\end{proposition}
}
The Laplace mechanism is a special instance of the exponential mechanism \shortonly{\cite{mcsherry07}} for real spaces $(\reals^{n}, \ell_{1})$.  
\begin{definition}[Laplace Mechanism]
Let $(\reals^{n}, \ell_{1})$ be the space of private data. The Laplace mechanism is defined as:
	\begin{align}
		Qu = u + V, \; \text{where } V \sim e^{ -\epsilon \|V\|_{1} }.
	\end{align}
\end{definition}

\begin{figure} \begin{center}
\includegraphics[width=.6\textwidth]{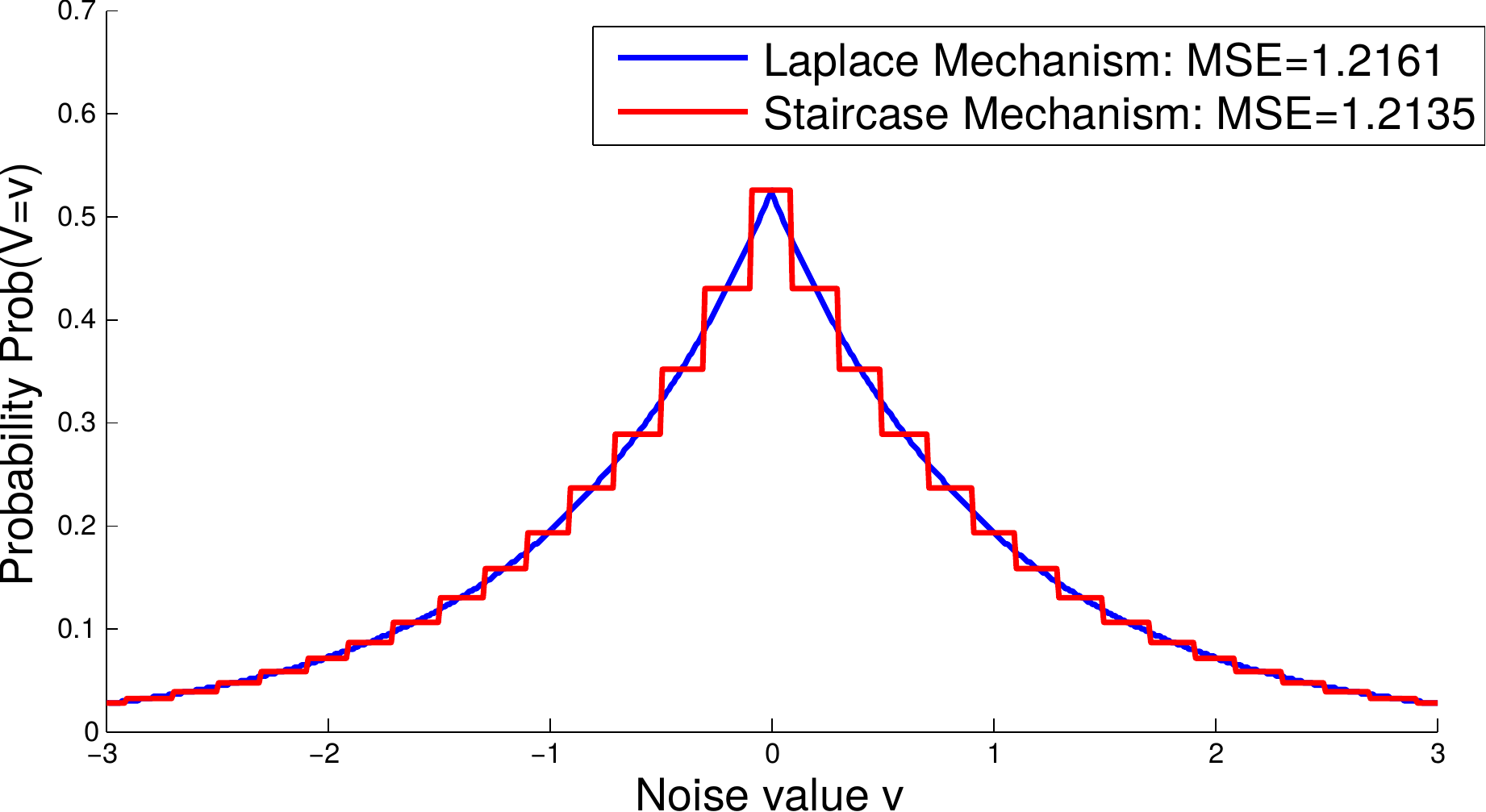}\caption{The staircase mechanism is the optimal $\epsilon$-differential private mechanism, whereas the Laplace mechanism is the optimal $\epsilon$-Lipschitz private mechanism. Therefore, the }  \label{fig:staircaseMechanism}
\end{center} \end{figure}

The Laplace mechanism can be shown to be $\epsilon$-differentially private. In general, however, the Laplace mechanism is suboptimal in the sense of minimum mean-squared error. 
For the single-dimensional case, the staircase mechanism~\cite{geng12} is the optimal $\epsilon$-differentially private mechanism; the mechanism which adds noise $V$ whose distribution is shown in Figure~\ref{fig:staircaseMechanism}. However, the Laplace mechanism is proven to be the optimal $\epsilon$-Lipschitz private mechanism in the sense of both minimum entropy~\cite{wang14} and minimum mean-squared error~\cite{koufogiannis15laplaceOptimality}, whereas the staircase mechanism fails to satisfy Lipschitz privacy due to its discontinuous probability density function.

\begin{theorem}[\cite{koufogiannis15laplaceOptimality} Optimality of Laplace] \label{thm:laplaceOptimalityL1}
Consider the $\epsilon$-Lipschitz private (in $(\reals^{n}, \ell_{1})$) mechanism $Q:\reals^{n}\rightarrow \Delta\left( \reals^{n} \right)$ of the form $Qu = u + V$, with $V\sim g(V) \in \Delta \left( \reals^{n} \right)$. Then, the Laplace mechanism that adds noise with density $l^{n}_{1}(v) = \left(\frac{\epsilon}{2}\right)^{n} e^{-\epsilon \|v\|_{1}}$ minimizes the mean-squared error. Namely, for any density $g$, we have:
\begin{align}
\expe\|Qu - u \|_{2}^{2} = \underset{V\sim g}{\expe} \|V\|^{2} \geq \underset{ V \sim l^{n}_{1} }{ \expe } \|V\|_{2}^{2} = \frac{2n}{\epsilon^{2}}.
\end{align}
\end{theorem}

\longonly{The optimal private mechanism characterizes the privacy-performance trade-off and is required for gradually releasing sensitive data. Thus, optimality of the Laplace mechanism in Theorem \ref{thm:laplaceOptimalityL1} is a key ingredient in our results and renders the problem tractable.}

\section{Gradual Release of Private Data} \label{sec:gradualPrivacy}
The problem of gradually releasing private data is now formulated. Initially, we focus on a single privacy level relaxation from to $\epsilon_{1}$ to $\epsilon_{2}$ and a single-dimensional space of private data $\mathcal{U}=\reals$. Subsections \ref{subsec:L1GradualPrivacy} and \ref{subsec:multiStep} present extensions to high-dimensional spaces and multiple rounds of privacy level relaxations, respectively.

Consider two privacy levels $\epsilon_{1}$ and $\epsilon_{2}$ with $\epsilon_{2}\geq\epsilon_{1}>0$. We wish to design a composite mechanism $Q_{\epsilon_{1} \to \epsilon_{2}}:\mathcal{U}\rightarrow \Delta\left( \mathcal{Y} \times \mathcal{Y} \right)$ that performs gradual release of data. The first and second coordinates respectively refer to the initial $\epsilon_{1}$-private and the subsequent $\epsilon_{2}$-private responses. In practice, given privacy levels $\epsilon_{1}$ and $\epsilon_{2}$ and an input $u\in\mathcal{U}$, we sample $(y_{1},y_{2})$ from the distribution $Q_{\epsilon_{1} \to \epsilon_{2}}u$. Initially, only coordinate $y_{1}$ is published satisfying $\epsilon_{1}$-privacy guarantees. Once privacy level is relaxed to $\epsilon_{2}$, response $y_{2}$ is released as a more accurate response of the \textit{same} query on the \textit{same} private data.

An adversary that wishes to infer the private input $u$ eventually has access to both responses $y_{1}$ and $y_{2}$. Therefore, the pair $(y_{1},y_{2})$ needs to satisfy $\epsilon_{2}$-privacy. On the other hand, an honest user wishes to maximize the accuracy of the response and, therefore, she is tempted to use an estimator $y_{M} = \theta(y_{1},y_{2})$ and infer a more accurate response $y_{M}$. In order to relieve honest users from any computational burden, we wish the best estimator to be as the truncation:
\begin{align} \label{eqn:truncateEstimator}
	\theta(y_{1},y_{2}) = y_{2}.
\end{align}

The composition theorem \cite{mcsherry07} provides a trivial, yet highly conservative, approach. Specifically, compositional rules imply that, if $y_{1}$ satisfies $\epsilon_{1}$-privacy and $(y_{1},y_{2})$ satisfies $\epsilon_{2}$-privacy, coordinate $y_{2}$ itself should be $(\epsilon_{2}-\epsilon_{1})$-private. In the extreme case that $\epsilon_{2}-\epsilon_{1} = \delta \ll 1$, response $y_{2}$ alone is expected to be $\delta$-private and, therefore, is highly corrupted by noise. This is unacceptable, since estimator \eqref{eqn:truncateEstimator} yields an even noisier response than the initial response $y_{1}$. Even if honest users are expected to compute more complex estimators than the truncation one in \eqref{eqn:truncateEstimator}, the approach dictated by composition theorem can still be unsatisfactory.

Specifically, consider the following two scenarios:
\begin{enumerate}
	\item \label{sce:twoStep} An $\epsilon_{1}$-private response $y_{1}$ is initially released. Once privacy level is relaxed from $\epsilon_{1}$ to $\epsilon_{2}$, an supplementary response $y_{2}$ is released.
	\item \label{sce:oneStep} No response is initially released. Response $\hat{y}_{2}$ is released as soon as the privacy level is relaxed to $\epsilon_{2}$.
\end{enumerate}
Then, there is no guarantee that the best estimator $\theta(y_{1},y_{2})$ in Scenario \ref{sce:twoStep} will match the accuracy of the response $\hat{y}_{2}$ in Scenario \ref{sce:oneStep}. An accuracy gap between the two scenarios would severely impact differential privacy. Specifically, the system designer needs to be strategic when choosing a privacy level. Differently stated, a market of private data based on composition theorems would exhibit \textit{friction}.

The key idea to overcome this friction is to introduce correlation between responses $y_{1}$ and $y_{2}$. In this work, we focus on Euclidean spaces $\mathcal{U}=\reals^{n}$ and mechanisms $Qu = u + V$ that approximate the identity query $q(u)=u$. Our main result states that a \textit{frictionless} market of private data is feasible and Scenarios \ref{sce:twoStep} and \ref{sce:oneStep} are equivalent. \longonly{This result has multi-fold implications:
\begin{itemize}
	\item A system designer is not required to be strategic with the choice of the privacy level. Specifically, she can initially under-estimate the required privacy level with $\epsilon_{1}$ and she can later fine-tune it to $\epsilon_{2}$ without hurting the accuracy of the final response.
	\item A privacy data market can exist and private data can be traded ``\textit{by the pound}''. An $\epsilon_{1}$-private response $y_{1}$ can be initially purchased. Next, a supplementary payment can be made in return for a privacy level relaxation to $\epsilon_{2}$ and a refined response $y_{2}$. The accuracy of the refined response $y_{2}$ is, then, unaffected by the initial transaction and is controlled only by the final privacy level $\epsilon_{2}$.
\end{itemize}
}

More concretely, given privacy levels $\epsilon_{1}$ and $\epsilon_{2}$ with $\epsilon_{2}>\epsilon_{1}$, we wish to design a composite mechanism $Q_{\epsilon_{1}\to\epsilon_{2}} : \mathcal{U}\to\mathcal{Y}\times\mathcal{Y}$ with the following properties:
\begin{enumerate}
	\item \label{req:perf1} The restriction of $Q_{\epsilon_{1}\to\epsilon_{2}}$ to the first coordinate should match the performance of the optimal $\epsilon_{1}$-private mechanism $Q_{\epsilon_{1}}$. More restrictively, the first coordinate of the composite mechanism $Q_{\epsilon_{1}\to\epsilon_{2}}$ should be distributed identically to the optimal $\epsilon_{1}$-private mechanism $Q_{\epsilon_{1}}$:
		\begin{align} \label{eqn:abstractGradReq:1}
			\mathbb{P}\left( Q_{\epsilon_{1}\to\epsilon_{2}}u \in \mathcal{S}\times\mathcal{Y} \right) = \mathbb{P}\left( Q_{\epsilon_{1}}u \in \mathcal{S} \right) , \: \forall u\in\mathcal{U} \text{ and } \mathcal{S}\subseteq\mathcal{Y}
		\end{align}
	\item \label{req:priv1} The restriction of $Q_{\epsilon_{1}\to\epsilon_{2}}$ to the first coordinate should be $\epsilon_{1}$-private. This property is imposed by constraint \ref{req:perf1}.
	\item \label{req:perf2} The restriction of $Q_{\epsilon_{1}\to\epsilon_{2}}$ to the second coordinate should match the performance of the optimal $\epsilon_{2}$-private mechanism $Q_{\epsilon_{2}}$. Similarly to the first coordinate, the second coordinate of the composite mechanism $Q_{\epsilon_{1}\to\epsilon_{2}}$ must be distributed identically to the optimal $\epsilon_{2}$-private mechanism $Q_{\epsilon_{2}}$:
		\begin{align} \label{eqn:abstractGradReq:2}
			\mathbb{P}\left( Q_{\epsilon_{1}\to\epsilon_{2}}u \in \mathcal{Y}\times\mathcal{S} \right) = \mathbb{P}\left( Q_{\epsilon_{2}}u \in \mathcal{S} \right) , \: \forall u\in\mathcal{U} \text{ and } \mathcal{S}\subseteq\mathcal{Y}
		\end{align}
	\item \label{req:priv2} Once both coordinates are published, $\epsilon_{2}$-privacy should be guaranteed. According to Lipschitz privacy, the requirement is stated as follows:
		\begin{align} \label{eqn:abstractGradReq:3}
			\mathbb{P}\left( Q_{\epsilon_{1}\to\epsilon_{2}}u\in \mathcal{S} \right) & \text{ is $\epsilon_{2}$-Lipschitz in $u$, for all } \mathcal{S}\subseteq\mathcal{Y}^{2}.
		\end{align}
\end{enumerate}

Equations \eqref{eqn:abstractGradReq:1} and \eqref{eqn:abstractGradReq:2} require knowledge of the optimal $\epsilon$-private mechanism. In general, computing the $\epsilon$-private mechanism that maximizes a reasonable performance criterion is still an open problem. Theorem \ref{thm:laplaceOptimalityL1} establish the optimality of the Laplace mechanism as the optimal private approximation of the identity query.

\subsection{Single-Dimensional Case}
Initially, we consider the single-dimensional case where $\mathcal{U}=\reals$ equipped with the absolute value. Theorem \ref{thm:laplaceOptimalityL1} establish the optimal $\epsilon$-private mechanism that is required by Equations \eqref{eqn:abstractGradReq:1} and \eqref{eqn:abstractGradReq:2}:
\begin{align} \label{eqn:gradPrivacy1D:1}
	Q_{\epsilon}u = u + V, \text{ where } V\sim e^{-\epsilon|V|}.
\end{align}
Mechanism \eqref{eqn:gradPrivacy1D:1} minimizes the mean-squared error from the identity query among all $\epsilon$-private mechanisms that use additive noise:
\begin{align}
	\underset{V\sim e^{-\epsilon |V|}}{\expe} (Q_{\epsilon}u - u)^{2}
\end{align}
Theorem \ref{thm:2Level1DGradualPrivacy} establishes the existence of a composite mechanism that relaxes privacy from $\epsilon_{1}$ to $\epsilon_{2}$ \textit{without any loss of performance}.

\begin{theorem} \label{thm:2Level1DGradualPrivacy}
Consider privacy levels $\epsilon_{1}$ and $\epsilon_{2}$ with $\epsilon_{2}\geq\epsilon_{1}>0$, and mechanisms of the form:
\begin{align}
Q_{1}u := u + V_{1} \text{ and } Q_{2}u := u + V_{2}, \text{ with } (V_{1},V_{2})\sim g\in\Delta\left(\reals^{2}\right).
\end{align}
Then, for density $l_{\epsilon_{1},\epsilon_{2}}$ with:
\begin{align} \label{eqn:twoLevelJointDistribution}
	l_{\epsilon_{1},\epsilon_{2}}(x,y) = \frac{ \epsilon_{1}^{2} }{ 2\epsilon_{2} } e^{-\epsilon_{2} |y|} \delta(x-y) + \frac{ \epsilon_{1} (\epsilon_{2}^{2}-\epsilon_{1}^{2}) }{ 4\epsilon_{2} } e^{-\epsilon_{1} |x-y| - \epsilon_{2} |y|},
\end{align}
where $\delta$ is the Dirac delta function, the following properties hold:
	\begin{enumerate}
		\item The mechanism $Q_{1}$ is $\epsilon_{1}$-private.
		\item The mechanism $Q_{1}$ is optimal, i.e. $Q_{1}$ minimizes the mean-squared error $\expe V_{1}^{2}$.
		\item The mechanism $\left( Q_{1}, Q_{2} \right)$ is $\epsilon_{2}$-private.
		\item The mechanism $Q_{2}$ is optimal, i.e. $Q_{2}$ minimizes the mean-squared error $\expe V_{2}^{2}$.
	\end{enumerate}
\end{theorem}
\longonly{
\begin{proof}
	Consider the mechanism $Q=(Q_{1},Q_{2})$ induced by the noise density \eqref{eqn:twoLevelJointDistribution}. We prove that this mechanism satisfies all the desired properties:
	\begin{enumerate}
		\item The first coordinate is Laplace-distributed with parameter $\frac{1}{\epsilon_{1}}$. For $x\geq0$, we get:
			\begin{align} \begin{split}
				\mathbb{P}(V_{1}=x) & = \int_{\reals} g(x,y) dy = \frac{ \epsilon_{1}^{2} }{ 2\epsilon_{2} } e^{-\epsilon_{2} x} + \frac{ \epsilon_{1} (\epsilon_{2}^{2}-\epsilon_{1}^{2}) }{ 4\epsilon_{2} } \int_{\reals} e^{-\epsilon_{1} |x-y| - \epsilon_{2} |y|} dy \\
					& = \frac{ \epsilon_{1}^{2} }{ 2\epsilon_{2} } e^{-\epsilon_{2} x} + \frac{ \epsilon_{1} (\epsilon_{2}^{2}-\epsilon_{1}^{2}) }{ 4\epsilon_{2} } \left( \int_{-\infty}^{0} e^{-\epsilon_{1}x + (\epsilon_{1}+\epsilon_{2})y} dy  +  \int_{0}^{x} e^{-\epsilon_{1}x - (\epsilon_{2} - \epsilon_{1})y} dy \right.  \\
					& \qquad \left. +  \int_{x}^{\infty} e^{\epsilon_{1}x - (\epsilon_{1}+\epsilon_{2})y} dy \right) \\
					& = \frac{ \epsilon_{1}^{2} }{ 2\epsilon_{2} } e^{-\epsilon_{2} x} + \frac{\epsilon_{1}(\epsilon_{2}-\epsilon_{1})}{4\epsilon_{2}} e^{-\epsilon_{1}x} \left. e^{(\epsilon_{1}+\epsilon_{2})y} \right|_{-\infty}^{0} - \frac{\epsilon_{1}(\epsilon_{1}+\epsilon_{2})}{4\epsilon_{2}} e^{-\epsilon_{1}x} \left. e^{-(\epsilon_{2}-\epsilon_{1})y} \right|_{0}^{x} \\
					& \qquad - \frac{\epsilon_{1}(\epsilon_{2}-\epsilon_{1})}{4\epsilon_{2}} \left. e^{-(\epsilon_{1}+\epsilon_{2})y} \right|_{x}^{\infty} \\
					& = \frac{\epsilon_{1}}{2} e^{-\epsilon_{1}x}
			\end{split} \end{align}
			The case $x\leq0$ follows from the symmetry $(x,y)\rightarrow(-x,-y)$. Therefore, the first coordinate is $\epsilon_{1}$-private and achieves optimal performance.
		\item The second coordinate is Laplace-distributed with parameter $\frac{1}{\epsilon_{2}}$. We have:
			\begin{align} \begin{split}
				\mathbb{P}(V_{2}=y) & = \int_{\reals} g(x,y) dx = \frac{ \epsilon_{1}^{2} }{ 2\epsilon_{2} } e^{-\epsilon_{2} |y|} + \frac{ \epsilon_{1} (\epsilon_{2}^{2}-\epsilon_{1}^{2}) }{ 4\epsilon_{2} } e^{-\epsilon_{2} |y|} \int_{\reals} e^{-\epsilon_{1}|x-y|} dx \\
					& = \frac{ \epsilon_{1}^{2} }{ 2\epsilon_{2} } e^{-\epsilon_{2} |y|} + \frac{ \epsilon_{1} (\epsilon_{2}^{2}-\epsilon_{1}^{2}) }{ 4\epsilon_{2} } e^{-\epsilon_{2} |y|} \int_{\reals} e^{-\epsilon_{1}|x|} dx \\
					& =  \frac{ \epsilon_{1}^{2} }{ 2\epsilon_{2} } e^{-\epsilon_{2} |y|} + \frac{\epsilon_{2}^{2}-\epsilon_{1}^{2}}{ 2\epsilon_{2} } e^{-\epsilon_{2} |y|} \\
					& = \frac{\epsilon_{2}}{2} e^{-\epsilon_{2} |y|}
				\end{split} \end{align}
			Thus, the second coordinate achieves optimal performance. 
		\item Lastly, we need to prove that the composite mechanism is $\epsilon_{2}$-private. We handle the delta part separately by defining $D = \{ x : \: (x,x)\in S \}$ for a measurable $S\subseteq\reals^{2}$. The probability of landing in set $S$ is:
			\begin{align} \begin{split}
				\mathbb{P}(Qu\in S) = \frac{\epsilon_{1}^{2}}{2\epsilon_{2}} \int_{D} e^{-\epsilon_{2} |x-u|} dx + \frac{\epsilon_{1}(\epsilon_{2}^{2}-\epsilon_{1}^{2})}{4\epsilon_{2}} \iint_{S} e^{-\epsilon_{1}|(x-u)-(y-u)|-\epsilon_{2}|y-u|} dx dy
			\end{split} \end{align}
			We take the derivative and use Fubini's theorem to exchange the derivative with the integral:
			\begin{align} \begin{split}
				\frac{d}{du}\mathbb{P}(Qu\in S) & = \frac{\epsilon_{1}^{2}}{2\epsilon_{2}} \int_{D} \epsilon_{2}\sgn(x-u) e^{-\epsilon_{2} |x-u|} dx \\
					& \qquad + \frac{\epsilon_{1}(\epsilon_{2}^{2}-\epsilon_{1}^{2})}{4\epsilon_{2}} \iint_{S} \epsilon_{2}\sgn(y-u) e^{-\epsilon_{1}|x-y|-\epsilon_{2}|y-u|} dx dy \Rightarrow \\
				\left| \frac{d}{du}\mathbb{P}(Qu\in S) \right| & \leq \frac{\epsilon_{1}^{2}}{2\epsilon_{2}} \int_{D} \epsilon_{2} e^{-\epsilon_{2} |x-u|} dx + \frac{\epsilon_{1}(\epsilon_{2}^{2}-\epsilon_{1}^{2})}{4\epsilon_{2}} \iint_{S} \epsilon_{2} e^{-\epsilon_{1}|(x-u)-(y-u)|-\epsilon_{2}|y-u|} dx dy \Rightarrow \\
				\left| \frac{d}{du}\mathbb{P}(Qu\in S) \right| & \leq \epsilon_{2} \mathbb{P}(Qu\in S) \Rightarrow \left| \frac{d}{du} \ln \mathbb{P}(Qu\in S) \right|  \leq \epsilon_{2}
			\end{split} \end{align}
	\end{enumerate}
	This completes the proof.
\end{proof}
}

\begin{figure} \begin{center}
\includegraphics[width=.9\textwidth]{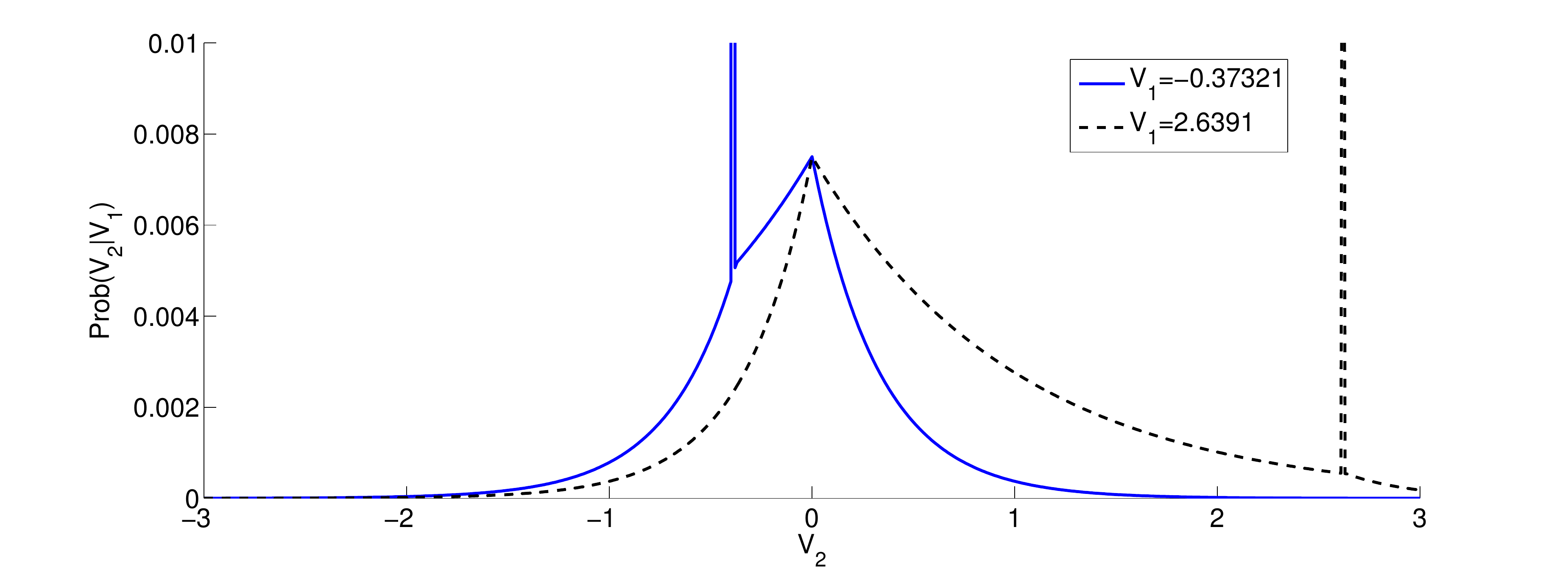}\caption{Gradual release of private data is performed in the following way. First, the $\epsilon_{1}$-private response $y_{1}=u+V_{1}$ is released, where $V_{1}\sim e^{-\epsilon|V_{1}|}$. Once privacy level is relaxed from $\epsilon_{1}=1$ to $\epsilon_{2}=2$, the supplementary response $y_{2}=u+V_{2}$ is released, where $V_{2}$ is distributed as shown above. The composite mechanism that releases $(y_{1},y_{2})$ is $\epsilon_{2}$-private and $V_{2}$ is optimally distributed.}  \label{fig:conditionalDistributions1D}
\end{center} \end{figure}

\shortonly{ The proof of Theorem \ref{thm:2Level1DGradualPrivacy} consists of checking that the expression for $l_{\epsilon_{1},\epsilon_{2}}$ satisfies the properties.}
\longonly{
\subsubsection{Single Round of Privacy Relaxation}
Theorem \ref{thm:2Level1DGradualPrivacy} achieves gradual release of sensitive data in two steps, first with $\epsilon_{1}$-privacy and, then, with $\epsilon_{2}$-privacy. In practice, Theorem \ref{thm:2Level1DGradualPrivacy} can be used as follows:
\begin{itemize}
	\item Given the private value $u\in\reals$, sample noise $V_{1}\sim e^{-\epsilon_{1}|V_{1}|}$ and release response $y_{1} = u + V_{1}$, which is optimal and respects $\epsilon_{1}$-privacy.
	\item Once privacy level is relaxed from $\epsilon_{1}$ to $\epsilon_{2}$, sample noise $V_{2}$ from the conditioned on $V_{1}$ distribution:
		\begin{align} \label{eqn:twoLevelConditionalDistribution}
			\prob( V_{2}=y | V_{1}=x ) = \frac{\epsilon_{1}}{\epsilon_{2}} e^{-(\epsilon_{2}-\epsilon_{1})|x|} \delta(y-x) + \frac{\epsilon_{2}^{2}-\epsilon_{1}^{2}}{2\epsilon_{2}} e^{ -\epsilon_{1}|y-x| - \epsilon_{2}|y| + \epsilon_{1}|x| },
		\end{align}
		and release response $y_{2} = u + V_{2}$. Distribution \eqref{eqn:twoLevelConditionalDistribution} is derived from the joint distribution \eqref{eqn:twoLevelJointDistribution} and ensures both that $(y_{1},y_{2})$ is $\epsilon_{2}$-private and that $V_{2}$ is optimally distributed.
\end{itemize}

Conditional distribution \eqref{eqn:twoLevelConditionalDistribution} is shown in Figure \ref{fig:conditionalDistributions1D}. Note that for $\epsilon_{2}=\epsilon_{1}$, Distribution \eqref{eqn:twoLevelConditionalDistribution} is reduced to a delta function:
\begin{align}
	\prob ( V_{2}=y | V_{1}=x ) = \delta(x-y).
\end{align}
In words, for $\epsilon_{2}=\epsilon_{1}$ no privacy relaxation effectively happens and, thus, no updated response is practically released. Moreover, for $\epsilon_{2}\to\infty$, a limiting argument shows that Distribution \ref{eqn:twoLevelConditionalDistribution} is reduced to:
\begin{align}
	\prob ( V_{2}=y | V_{1}=x ) = \delta(y).
\end{align}
Practically, letting $\epsilon_{2}\to\infty$ cancel any privacy constraints and the exact value of private data $u$ can be released $y_{2} = u$. For general values of $\epsilon_{1}$ and $\epsilon_{2}$, Pearson's correlation coefficient decreases for more aggressive privacy relations, $\rho_{V_{1},V_{2}} = \frac{\epsilon_{1}}{\epsilon_{2}}$. Algorithm \ref{alg:2LevelGradPrivacy} provides a simple and efficient way to sample $V_{2}$ given $V_{1}$.
}

\longonly{
\begin{algorithm}
	\begin{algorithmic}
		\Require{Privacy levels $\epsilon_{1}$ and $\epsilon_{2}$, such that $\epsilon_{2}>\epsilon_{1}>0$, and noise sample $x$.}
		\Function{RelaxPrivacy}{$x,\epsilon_{1},\epsilon_{2}$}
			\Switch{randomly}
				\Case{ with probability $\frac{\epsilon_{1}}{\epsilon_{2}} e^{-(\epsilon_{2}-\epsilon_{1})|x|}$: }
					\State return $y=x$.
				\EndCase
				\Case{ with probability $\frac{\epsilon_{2}-\epsilon_{1}}{2\epsilon_{2}}$: }
					\State draw $z\sim \begin{cases} e^{(\epsilon_{1}+\epsilon_{2})z}, \: & \text{ for } z\leq0 \\ 0, \: & \text{ otherwise.} \end{cases} $
					\State return $y = \sgn(x)z$.
				\EndCase
				\Case{ with probability $\frac{\epsilon_{1}+\epsilon_{2}}{2\epsilon_{2}} \left( 1 - e^{-(\epsilon_{2}-\epsilon_{1})|x|} \right)$: }
					\State draw $z\sim \begin{cases} e^{-(\epsilon_{2}-\epsilon_{1})z}, \: & \text{ for } 0\leq z\leq|x| \\ 0, \: & \text{ otherwise.} 		\end{cases} $
					\State $y = \sgn(x)z$.
				\EndCase
				\Case{ with probability $\frac{\epsilon_{2}-\epsilon_{1}}{2\epsilon_{2}} e^{-(\epsilon_{2}-\epsilon_{1})|x|}$: }
					\State draw $z\sim \begin{cases} e^{-(\epsilon_{1}+\epsilon_{2})z}, \: & \text{ for } z\geq|x| \\ 0, \: & \text{ otherwise.} \end{cases} $
					\State return $y = \sgn(x)z$.
				\EndCase 
			\EndSwitch
		\EndFunction
	\end{algorithmic}
	\caption{Sampling from Distribution (\ref{eqn:twoLevelConditionalDistribution}) for the second noise sample $V_{2}=y$ given the first noise sample $V_{1}=x$ can be efficiently performed.} \label{alg:2LevelGradPrivacy}
\end{algorithm}
}

\longonly{
\subsubsection{Single Round of Privacy Tightening}
Tightening the privacy level is impossible, since it implies revoking already released data. Nonetheless, generating a more private version of the same data is still useful in cases such as \textit{private data trading}. In that case, distribution \eqref{eqn:twoLevelJointDistribution} can be sampled in the opposite direction. Specifically, noise $V_{2}$ is initially sampled, $V_{2} \sim e^{-\epsilon_{2}|V_{2}|}$, and the $\epsilon_{2}$-private response $y_{2} = u+V_{2}$ is released. Next, private data $u$ is traded to a \textit{different agent} under the stronger $\epsilon_{1}$-privacy guarantees. Noise sample $V_{1}$ is drawn from distribution
\begin{align} \label{eqn:backwardConditionalDistribution}
\prob( V_{1} = x | V_{2} = y ) =  \left( \frac{\epsilon_{1}}{\epsilon_{2}} \right)^{2}  \delta(x-y) + \left( 1 - \left( \frac{\epsilon_{1}}{\epsilon_{2}} \right)^{2} \right) \frac{ \epsilon_{1} }{2} e^{-\epsilon_{1} |x-y|},
\end{align}
and the $\epsilon_{1}$-private response $y_{1} = u + V_{1}$ is released. Remarkably, response $y_{1}$ can be generated conditioning only on $y_{2}$:
\begin{align}
y_{2} = y_{1} + V_{2\to1},
\end{align}
where $V_{2\to1} = V_{1} - V_{2}$ is independent of $V_{2}$, $V_{2\to1} \bot V_{2}$. In words, tightening privacy under the Laplace mechanism does not require access to the original data $u$ and can be performed by an agent other than the private data owner. Theorem \ref{thm:postprocessing} suggests that the randomized post-processing $y_{1} = y_{2} + V_{2\to1}$ of the $\epsilon_{2}$-private response $y_{2}$ is at least $\epsilon_{2}$-private. For $V_{2\to1}$ given by distribution \eqref{eqn:backwardConditionalDistribution}, this tightening of privacy level is precisely quantified, i.e. $\epsilon_{2}\to\epsilon_{1}$. Recall that our results are \textit{tight}; no excessive accuracy is sacrificed in the process.
}

\subsection{High-Dimensional Case} \label{subsec:L1GradualPrivacy}
Theorem \ref{thm:2Level1DGradualPrivacy} can be generalized for the case that the space of private data is Euclidean $\reals^{n}$ equipped with the $\ell_{1}$-norm. Theorem \ref{thm:laplaceOptimalityL1} establishes that the Laplace mechanism:
\begin{align} \label{eqn:gradPrivacy1D:2}
	Q_{\epsilon}u = u + V, \text{ where } V\sim e^{-\epsilon \|V\|_{1}}.
\end{align}
minimizes the mean-squared error from the identity query among all $\epsilon$-private mechanisms that use additive noise $V\in\reals^{n}$:
\begin{align}
	\underset{ V\sim e^{-\epsilon \|V\|_{1}} }{\expe} \|Q_{\epsilon}u - u\|_{2}^{2}.
\end{align}

Theorem \ref{thm:laplaceOptimalityL1} shows that each coordinate of $V$ is independently sampled. This observation implies that Theorem \ref{thm:2Level1DGradualPrivacy} can be applied to $n$ dimensions independently.

\begin{theorem} \label{thm:L1GradualMechanism}
Consider privacy levels $\epsilon_{1}$, $\epsilon_{2}$ with $\epsilon_{2}>\epsilon_{1}>0$. Let $Q_{1}$ be an $\epsilon_{1}$-private mechanism and $Q_{2}$ an $\epsilon_{2}$-private mechanism of the form:
\begin{align}
	Q_{1}u := u + V_{1} \text{ and } Q_{2}u := u + V_{2} \text{ with } (V_{1},V_{2})\sim g\in\Delta\left(\mathbb{R}^{2n}\right),
\end{align}
where $u \in \mathbb{R}_{l_{1}}^{n}$
Then, gradual release of sensitive data $u$ from $\epsilon_{1}$ to $\epsilon_{2}$ is achieved by the probability distribution $l_{\epsilon_{1},\epsilon_{2}}^{n}$:
\begin{align}
	l_{\epsilon_{1},\epsilon_{2}}^{n}( V_{1}, V_{2} ) = \prod_{i=1}^{n}  l_{\epsilon_{1},\epsilon_{2}}( V_{1}^{(i)}, V_{2}^{(i)}),
\end{align}
where $V_{i} = [V_{i}^{(1)},\ldots,V_{i}^{(n)}]$, $i=1,2$. Namely:
\begin{itemize}
	\item Mechanism $Q_{1}$ is $\epsilon_{1}$-private and optimal.
	\item Mechanism $Q_{2}$ is the optimal $\epsilon_{2}$-private mechanism.
	\item Mechanism $(Q_{1},Q_{2})$ is $\epsilon_{2}$-private.
\end{itemize}
\end{theorem}
\longonly{
\begin{proof}
Let $[x^{(1)},\ldots,x^{(n)}]$ denote the coordinates of a vector $x\in\reals^{n}$. The desired probability distribution is defined by independently sampling each coordinate using Theorem \ref{thm:2Level1DGradualPrivacy}. Let:
\begin{align}
l_{\epsilon_{1},\epsilon_{2}}^{n}(x,y) = g(x,y) = \prod_{i=1}^{n} l_{\epsilon_{1},\epsilon_{2}}(x^{(i)},y^{(i)}),
\end{align}
The probability distribution satisfies the required marginal distributions:
\begin{align*}
\int_{\mathbb{R}^{n}} g(x,y) d^{n}y = \left(\frac{\epsilon_{1}}{2}\right)^{n} e^{-\epsilon_{1} \|x\|_{1}} \quad \text{and} \quad \int_{\mathbb{R}^{n}} g(x,y) d^{n}x = \left(\frac{\epsilon_{2}}{2}\right)^{n} e^{-\epsilon_{2} \|y\|_{1}}
\end{align*}
Moreover, it satisfies $\epsilon_{2}$-privacy constraints:
\begin{align*} \begin{split}
	& \quad \left\| \nabla_{u} \ln \prob( Q_{1}u = z_{1} \text{ and } Q_{2}u = z_{2} ) \right\|_{\infty} \\
	& = \left\| \nabla_{u} \ln l_{\epsilon_{1},\epsilon_{2}}^{n}(z_{1}-u,z_{2}-u) \right\|_{\infty} \\
	& = \max_{1\leq i\leq n} \left| \frac{\partial}{\partial u_{i}} \ln l_{\epsilon_{1},\epsilon_{2}}^{n}(z_{1}-u,z_{2}-u) \right| \\
	& = \max_{1\leq i\leq n} \left| \frac{\partial}{\partial u_{i}} \ln l_{\epsilon_{1},\epsilon_{2}}(z_{1}^{(i)}-u^{(i)},z_{2}^{(i)}-u^{(i)}) \right| \\
	& \leq \max_{1\leq i\leq n} \epsilon_{2} = \epsilon_{2},
\end{split} \end{align*}
where in the last line we used the fact that $l_{\epsilon_{1},\epsilon_{2}}$ is $\epsilon_{2}$-private. This completes the proof.
\end{proof}
}

\shortonly{The proof of Theorem~\ref{thm:L1GradualMechanism} component-wisely invokes Theorem~\ref{thm:2Level1DGradualPrivacy} and confirms that distribution $l_{\epsilon_{1},\epsilon_{2}}^{n}$ possesses the desired properties.}

\subsection{Multiple Privacy Relaxations} \label{subsec:multiStep}
Theorems \ref{thm:2Level1DGradualPrivacy} and \ref{thm:L1GradualMechanism} perform privacy relaxation from $\epsilon_{1}$ to $\epsilon_{2}$. However, the privacy level is possibly updated multiple times. Theorem \ref{thm:multiStepPrivacy} handles the case where the privacy level is successively relaxed from $\epsilon_{1}$ to $\epsilon_{2}$, to $\epsilon_{3}$, until $\epsilon_{m}$. Specifically, Theorem \ref{thm:multiStepPrivacy} enables the use of Theorem \ref{thm:2Level1DGradualPrivacy} multiple times while relaxing privacy level from $\epsilon_{i}$ to $\epsilon_{i+1}$ for $i\in\{1,\ldots,m-1\}$. We call this statement the \textit{Markov property} of the Laplace mechanism.

\begin{theorem} \label{thm:multiStepPrivacy}
Consider $m$ privacy levels $\{\epsilon_{i}\}_{i=1}^{m}$ with $0<\epsilon_{1}<\cdots<\epsilon_{m}$ and  mechanisms $Q_{i}$ of the form:
\begin{align}
Q_{i}u = u + V_{i}, \text{ with } (V_{1},\ldots,V_{m}) \sim g \in \Delta\left(\mathbb{R}^{m}\right).
\end{align}
Consider the distribution $g=l_{\epsilon_{1},\ldots,\epsilon_{m}}$, with:
\begin{align} \label{eqn:multiStepDistribution}
	l_{\epsilon_{1},\ldots,\epsilon_{m}}(v_{1},\ldots,v_{m}) = l_{\epsilon_{1}}(v_{1}) \prod_{i=1}^{m-1} \frac{l_{\epsilon_{i},\epsilon_{i+1}}(v_{i},v_{i+1})}{l_{\epsilon_{i}}(v_{i})},
\end{align}
where $l_{\epsilon}(v) = \frac{\epsilon}{2} e^{-\epsilon |v|}$. Then, distribution $l_{\epsilon_{1},\ldots,\epsilon_{m}}$ has the following properties:
\begin{enumerate}
	\item Each prefix mechanism $\left(Q_{1}, \ldots, Q_{i}\right)$ is $\epsilon_{i}$-private, for $i\in\{1,\ldots,m\}$.
	\item Each mechanism $Q_{i}$ is the optimal $\epsilon_{i}$-private mechanism, i.e. it minimizes the mean-squared error $\expe V_{i}^{2}$.
\end{enumerate}
\end{theorem}

\longonly{
\begin{proof}
The proof uses induction on $m$. The case $m=2$ is handled by Theorem \ref{thm:2Level1DGradualPrivacy}. For brevity, we prove the statement for $m=3$. Let $f(x,y)=l_{\epsilon_{1},\epsilon_{2}}(x,y)$ and $g(y,z)=l_{\epsilon_{2},\epsilon_{3}}(y,z)$. Consider the joint probability $l_{\epsilon_{1},\epsilon_{2},\epsilon_{3}}$:
\begin{align} \begin{split}
h(x,y,z) = l_{\epsilon_{1},\epsilon_{2},\epsilon_{3}}(x,y,z) = \frac{ f(x,y) g(y,z) }{ l_{\epsilon_{2}}(y) },
\end{split} \end{align}
where $l_{\epsilon_{2}}(y) = \frac{\epsilon_{2}}{2} e^{-\epsilon_{2} |y|}$. Measure $h$ possesses all the properties that perform gradual release of private data:
\begin{itemize}
	\item All marginal distributions of measure $h$ are Laplace with parameters $\frac{1}{\epsilon_{1}}$, $\frac{1}{\epsilon_{2}}$, and $\frac{1}{\epsilon_{3}}$, respectively:
		\begin{align*}
			\int_{\mathbb{R}} \int_{\mathbb{R}} h(x,y,z) dy dz = l_{\epsilon_{1}}(x), \quad \int_{\mathbb{R}} \int_{\mathbb{R}} h(x,y,z) dx dz = l_{\epsilon_{2}}(y), \text{ and } \int_{\mathbb{R}} \int_{\mathbb{R}} h(x,y,z) dx dy = l_{\epsilon_{3}}(z).
		\end{align*}
	\item Mechanism $Q_{1}$ is $\epsilon_{1}$-private since $V_{1}$ is Laplace-distributed with parameter $\frac{1}{\epsilon_{1}}$.
	\item Mechanism $(Q_{1},Q_{2})$ is $\epsilon_{2}$-private. Margining out $V_{3}$ shows that $(V_{1},V_{2})\sim l_{\epsilon_{1},\epsilon_{2}}$, which guarantees $\epsilon_{2}$-privacy according to Theorem \ref{thm:2Level1DGradualPrivacy}.
	\item Mechanism $(Q_{1},Q_{2},Q_{3})$ is $\epsilon_{3}$-private. It holds that:
		\begin{align} \begin{split}
			& \quad \left| \frac{\partial}{\partial u} \prob( Q_{1}u=\psi_{1}, \; Q_{2}u=\psi_{2}, \text{ and } Q_{3}u=\psi_{3} ) \right| \\
			& = \left| \frac{\partial}{\partial u} h( \psi_{1}-u, \psi_{2}-u, \psi_{3}-u ) \right| \\
			& = \left. \left| \frac{\partial h(x,y,z)}{\partial x} + \frac{\partial h(x,y,z)}{\partial y} + \frac{\partial h(x,y,z)}{\partial z} \right| \right|_{ \substack{ x=\psi_{1}-u, \\ y=\psi_{2}-u, \\ z=\psi_{3}-u} }
		\end{split} \end{align}
		Algebraic manipulation of the last expression establishes the result:
		\begin{align} \begin{split}
			\left| \frac{\partial h}{\partial x} + \frac{\partial h}{\partial y} + \frac{\partial h}{\partial z} \right| & = \left| \frac{f_{x}g}{l_{\epsilon_{2}}} + \frac{f_{y}g}{l_{\epsilon_{2}}} + \frac{fg_{y}}{l_{\epsilon_{2}}} - l_{\epsilon_{2}} \frac{fg}{l_{\epsilon_{2}^{2}}} + \frac{fg_{z}}{l_{\epsilon_{2}}} \right| \\
			& = \left| - \sgn(y)\epsilon_{2}\frac{fg}{l_{\epsilon_{2}}} - \sgn(z)\epsilon_{3}\frac{fg}{l_{\epsilon_{2}}} + \sgn(y)\epsilon_{2}\frac{fg}{l_{\epsilon_{2}}} \right| \\
			& = \left|  - \sgn(z)\epsilon_{3}\frac{fg}{l_{\epsilon_{2}}} \right| \\
		& = \epsilon_{3}h,
		\end{split} \end{align}
\end{itemize}
where we used the properties $l'_{\epsilon_{2}} = -\sgn(y) \epsilon_{2} l_{\epsilon_{2}}$, $f_{x}+f_{y} = -\sgn(y) \epsilon_{2} f$, and $g_{y}+g_{z} = -\sgn(z) \epsilon_{3} g$, where the last two identities were derived in the proof of Theorem \ref{thm:2Level1DGradualPrivacy}.
\end{proof}
}
\shortonly{Proof of Theorem~\ref{thm:multiStepPrivacy} uses induction on the number of privacy levels $m$ and Theorem~\ref{thm:2Level1DGradualPrivacy} is invoked in the inductive step.}
Additionally, performing multiple rounds of privacy relaxations can be performed in the context of Theorem \ref{thm:L1GradualMechanism} is possible. In that case, Theorem \ref{thm:multiStepPrivacy} is independently applied to each component.

\subsubsection{Multiple Rounds of Privacy Relaxation}
Theorem \ref{thm:multiStepPrivacy} states that it is possible to repeatedly use Theorem \ref{thm:2Level1DGradualPrivacy} to perform multiple privacy level relaxations. An intuitive proof of Theorem \ref{thm:multiStepPrivacy} can be constructed by considering Scenarios \ref{sce:twoStep} and \ref{sce:oneStep} introduced in the beginning of Section \ref{sec:gradualPrivacy}. Specifically, Theorem \ref{thm:2Level1DGradualPrivacy} constructs a \textit{coupling} such that Scenario \ref{sce:twoStep} replicates Scenario \ref{sce:oneStep}. Therefore, once the first round of privacy relaxation $\epsilon_{1}\to\epsilon_{2}$  occurs, the two scenarios are indistinguishable. The second round of privacy relaxation $\epsilon_{2}\to\epsilon_{3}$ is performed by starting at the first step of Scenario \ref{sce:twoStep}.

In practice, Theorem \ref{thm:multiStepPrivacy} allows for an efficient implementation of an arbitrary number or privacy relaxation rounds $\epsilon_{1} \to \epsilon_{2} \to \ldots \to \epsilon_{m}$. In particular, only the most recent privacy level $\epsilon_{i}$ and noise sample $V_{i}$ need to be stored in memory. Sampling for $V_{i+1}$ depends only on current privacy level $\epsilon_{i}$, current noise sample $V_{i}$ and next privacy level $\epsilon_{i+1}$. Past privacy levels $\{\epsilon_{j}\}_{j<i}$, past noise samples $\{V_{j}\}_{j<i}$, and future privacy levels $\{\epsilon_{j}\}_{j>i+1}$ are not needed.

\begin{figure} \begin{center}
\includegraphics[width=.8\linewidth]{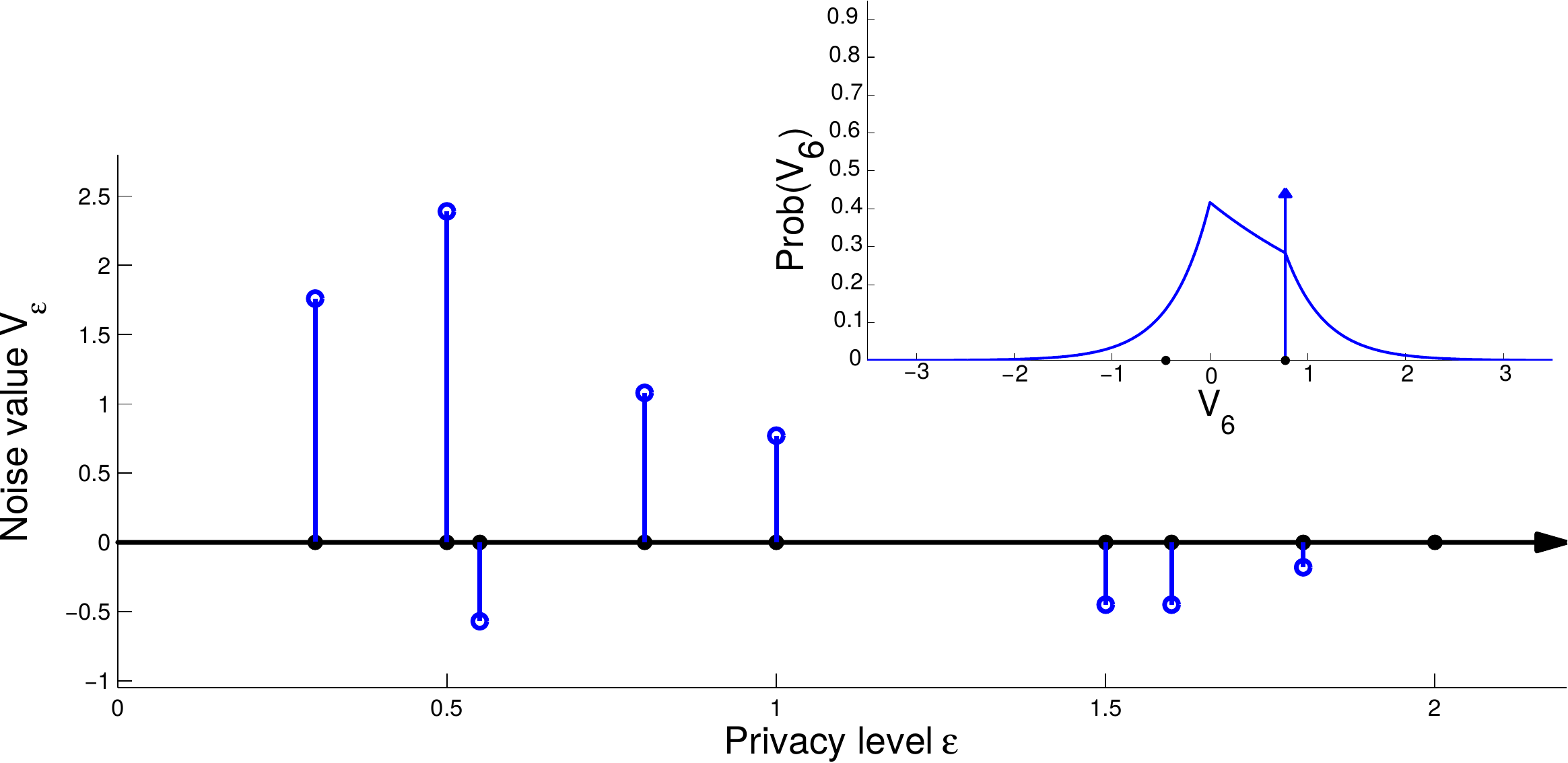}
\caption{Privacy level can be repeatedly relaxed. For each round of relaxation $\epsilon_{i} \to \epsilon_{i+1}$, the distribution of the next noise sample $V_{i+1}$ depends only on the last noise sample $V_{i}$. Past noise samples $\{V_{j}\}_{j<i}$ can be discarded from memory, thus, there is no complexity incurred from repeatedly relaxing privacy level.} \label{fig:multipleRounds}
\end{center} \end{figure}

\longonly{
\subsection{A Private Stochastic Process}
Theorems \ref{thm:2Level1DGradualPrivacy} and \ref{thm:multiStepPrivacy} offer a novel dimension to the Laplace mechanism. Specifically, these results establish a real-valued stochastic process $\{ V_{\epsilon} : \epsilon>0 \} $. Sampling from the process $\{V_{\epsilon}\}_{\epsilon>0}$ performs gradual release of sensitive data for the continuum of privacy levels $(0,\infty)$. Consider the mechanisms $Q_{\epsilon}$ that respond with $Q_{\epsilon}u = y_{\epsilon} =  u + V_{\epsilon}$. Then:
\begin{itemize}
	\item $V_{\epsilon}$ is optimally distributed, i.e. Laplace-distributed with parameter $\frac{1}{\epsilon}$
	\item Any $\epsilon$-truncated response $\{ y_{\sigma} \}_{ \sigma\in(0,\epsilon] }$ is $\epsilon$-private.
\end{itemize}

Samples of the process $\{V_{\epsilon}\}_{\epsilon>0}$ are plotted in Figure \ref{fig:privateProcess1D}. This process features properties that allow efficient sampling:
\begin{itemize}
	\item It is \textit{Markov}, $V_{s} \bot  V_{t} | V_{q}$, for $s<q<t$. Thus, a sample of the process $V_{\epsilon}$ over an interval $[\epsilon_{1},\epsilon_{2}]$ can be extended to $[\epsilon_{1},\epsilon_{3}]$, for $\epsilon_{3}>\epsilon_{2}$.
	\item It is \textit{lazy}, i.e. $V_{\epsilon+\delta} = V_{\epsilon}$ with high probability, for $\delta\ll1$. Therefore, A sample of the process $\{V_{\epsilon}\}_{\epsilon_{1} \leq \epsilon \leq \epsilon_{2}}$ can be efficiently stored; only a finite (random) number $m$ of points $(\epsilon_{i}, V_{\epsilon_{i}})_{i=1}^{m}$ where jumps occur need to be stored for \textit{exact} re-construction of the process.
\end{itemize}
}

\longonly{

\section{Applications}


\subsection{Crowdsourcing Statistics with RAPPOR} \label{sec:rappor}
Theorems \ref{thm:2Level1DGradualPrivacy} and \ref{thm:multiStepPrivacy} perform gradual release of private data by releasing responses that approximate the identity query $q(u) = u$. In practice, however, the end-user of private data is interested in more expressive queries $q$. The spectrum of such queries vastly varies. Examples include the mean value $\frac{1}{n} \sum_{i=1}^{n}u_{i}$ of a collection of private data $u_{1},\ldots,u_{n}$, and solutions to optimization problems \cite{han2014differentially}. Our results are directly applicable to a broad family of queries which are approximated by private mechanisms built around the Laplace mechanism. Specifically, consider mechanisms based on the Laplace mechanism and have the form shown in Figure \ref{fig:laplaceBasedMechanism}. The database of private data is initially preprocessed and, then, additive Laplace-distributed noise is used. The result is post-processed in order to maximize the accuracy of the response. Informally stated:

\begin{corollary}
Let $(\mathcal{U},d)$ be a metric space of sensitive data, $\mathcal{Y}$ be a set of responses, and $\epsilon>0$ be a privacy level. Let
	\begin{itemize}
		\item $F: \mathcal{U} \to \Delta \left( \reals^{n} \right)$ be a preprocessing step with sensitivity $\beta$ that is invariant of $\epsilon$,
		\item $\mathcal{L}_{\epsilon}: \reals^{n} \to \Delta \left( \reals^{n} \right)$ be the Laplace mechanism with parameter $\epsilon$:
			\begin{align}
				\mathcal{L}_{\epsilon}u = u + V, \text{ where } V \sim e^{- \frac{\epsilon}{\beta} \|V\|_{1}},
			\end{align}
		\item $G_{\epsilon}: \reals^{n} \to \Delta \left( \mathcal{Y} \right)$ be a post-processing step.
	\end{itemize}
Consider the $\epsilon$-private mechanism
\begin{align}
	G \circ \mathcal{L} \circ F : \mathcal{U} \to \Delta \left( \mathcal{Y} \right).
\end{align}
Then, there exists a composite mechanism that performs gradual release of sensitive data $u \in \mathcal{U}$.
\end{corollary}

Thus, our results are directly applicable to a existing privacy-aware mechanisms in e.g. smart grids \cite{koufogiannis14}, \cite{erlingsson2014rappor}, and user's reports \cite{erlingsson2014rappor}. On the other hand, applying our results is not yet possible for mechanisms that do not fulfill this assumption, such as privately solving optimization problems with stochastic gradient descent \cite{han2014differentially}.

\begin{figure} \begin{center}
\includegraphics[width=.9\linewidth]{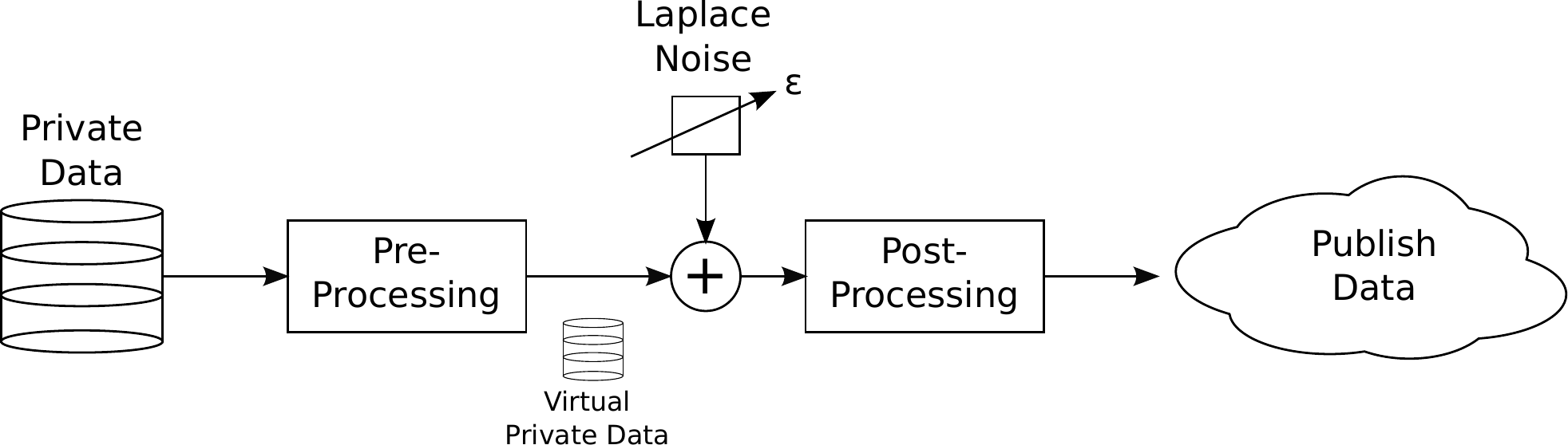}
\caption{User $1$ wants to share his sensitive data, such as his date of birth, in the a social network. Although, user $1$ has no privacy concerns when sharing this information with his close friends $2$ and $3$, he has gradually increasing privacy issues for other members of the network. Specifically, a group $A$ of distant users should not be able to collude and extract more information than what it is intended.} \label{fig:laplaceBasedMechanism}
\end{center} \end{figure}

In particular, Google's RAPPOR \cite{erlingsson2014rappor} is a mechanism that collects private data from multiple users for ``crowdsourcing statistics'' and can be expressed in terms of the Laplace mechanism. RAPPOR collects personal information from users such as the software features they use and the URLs they visited, and provides statistics of this information over a population of users. Algorithmically, a Bloom filter $B$ is applied of size $k$ is applied to each user's private data $u$:
\begin{align}
B: \mathcal{U} \to \{0,1\}^{k}, \quad y = [y_{1}, \ldots, y_{k}] = B(u),
\end{align}
where $\mathcal{U}$ is the space of private data, in particular, the set of all strings. Next, each bit $y_{i}$ is perturbed with probability $f$ and the result is memoised:
\begin{align} \begin{split} \label{eqn:rappor:1}
f: \{0,1\}^{k} \rightarrow \{0,1\}^{k}, \quad z = [z_{1}, \ldots, z_{k}] = f(y), 
\text{ where } z_{i} = 	\begin{cases}
							0, \quad & \text{w.p. } \frac{1}{2} \alpha, \\
							1, \quad & \text{w.p. } \frac{1}{2} \alpha, \\
							y_{i}, \quad & \text{w.p. } 1-c_{1},
						\end{cases}
\end{split} \end{align}
where ``w.p.'' stands for ``with probability'' and $\alpha\in[0,1]$ is a parameter. Finally, RAPPOR applies another perturbation each time a report is communicated to the server. This perturbation is equivalent to the map \eqref{eqn:rappor:1} but differently parametrized:
\begin{align} \begin{split} \label{eqn:rappor:2}
g: \{0,1\}^{k} \rightarrow \{0,1\}^{k}, \quad w = [w_{1}, \ldots, w_{k}] = f(z), 
\text{ where } \prob( w_{i}=1 ) =	\begin{cases}
										\beta, \quad & \text{if } z_{i} = 1, \\
										\gamma, \quad & \text{if } z_{i} = 0,
									\end{cases}
\end{split} \end{align}
where $\beta,\gamma\in[0,1]$ are parameters. RAPPOR's differential privacy guarantees relax (increased $\epsilon$) for small values of $\alpha$ and $\gamma$, and large values of $\beta$.

An important limitation of RAPPOR is that parameters $\alpha$, $\beta$, and $\gamma$ are forever fixed. However, there are reasons that require the ability to update these values in a way that the privacy is relaxed and the accuracy is increased:
\begin{itemize}
	\item Due to the non-trivial algorithm of decoding the reports, a tight accuracy-analysis is not possible. Instead, the accuracy of the system is evaluated once the system is bootstrapped.\footnote{Even in that case, estimating the actual accuracy can be challenging since it should be performed in a differential private way.} Our results makes it possible to initialize the parameters with tight values $\alpha\to1$, $\beta\to.5$, $\gamma\to.5$, and subsequently relax the parameters until a desired accuracy is achieved.
	\item Once a process or URL is suspected as malicious, the server would be interested in relaxing the privacy level and performing more accurate analysis of the potential threat. Once such a threat is identified, our result allows users to gradually relax their privacy parameters and the server can more confidently explore the potential threat.
\end{itemize}

In order to apply Theorems \ref{thm:2Level1DGradualPrivacy} and \ref{thm:multiStepPrivacy} to RAPPOR, we express the randomized maps \eqref{eqn:rappor:1} and \eqref{eqn:rappor:2} using the Laplace mechanism. Specifically, consider the functions $\bar{f}$ and $\bar{g}$ that add Laplace noise and project the result to $\{0,1\}$:
\begin{align}
\bar{f}(\psi) & = \left[ \psi + V_{f} > \frac{1}{2} \right], \quad \text{where } V_{f} \sim \text{Lap}\left( \frac{1}{-2\ln \alpha} \right), \\
\bar{g}(\zeta) & = \left[ \zeta + V_{g} > \frac{ \ln(2\gamma) }{ \ln\left( 4\gamma(1-\beta) \right) } \right], \quad \text{where } V_{g} \sim \text{Lap}\left( \frac{1}{- \ln \left( 4\beta(1-\gamma) \right)} \right),
\end{align}
where $\psi,\zeta\in\{0,1\}$, $\text{Lap}(b)$ is the Laplace distribution with parameter $b$, and $[\mathtt{expr}]\in\{0,1\}$ is $1$ if, and only if,  $\mathtt{expr}$ is true. Note that functions $\bar{f}$ and $\bar{g}$ have the structure of Figure \ref{fig:laplaceBasedMechanism}. Moreover, it can be shown that $\bar{f}$ and $\bar{g}$ applied component-wise to $y$ and $z$ are reformulations of the maps $f$ and $g$. Therefore, privacy level relaxation is achieved by sampling noises $V_{f}$ and $V_{g}$ as suggested by our results.

\todo{I ll need to be more careful with the composition of the two mechanisms. Maybe, I should express the composition $g\circ f$ as a Laplace mechanism instead.}

\subsection{Privacy in Social Networks} \label{sec:socialNetwork}
The context of social networks provides another setting where gradually releasing private data is critical. Consider a social network as a graph $G=(V,E)$, where $V$ is the set of users and $E$ the set of friendships between them. Each user owns a set of sensitive data that can include the date of birth, the number of friends and the city he currently resides. In the realm of social media, user's privacy concerns scale with the distance to other users of the network. Specifically, an individual is willing to share his private data with his close friends without any privacy guarantees, is skeptical about sharing this information with friends of his friends, and is alarmed to release his sensitive data over the entire social network. Therefore, an individual $i$ chooses a different privacy level $\epsilon_{j}$ for each user $j\in V$ as a decreasing function between users $i$ and $j$:
\begin{align}
	\epsilon_{j} = \frac{1}{d(i,j)},
\end{align}
where $d$ is a distance measure, e.g. the length of the shortest path between nodes $i$ and $j$. Then, user $i$ could generate an $\epsilon_{j}$-private response $y_{j}$ independently for each member $j$ of the network. However, more private information than desired is leaked. Specifically, consider the part of the social network shown in Figure \ref{fig:socialNetwork}, where user $i=1$ wishes to share his sensitive data $u$, such as her date of birth. Then, consider a group $A\subseteq V$ of users residing far away from user $1$ such that the privacy budget $\epsilon_{j}$ allocated by user $i$ to each member $j$ of the group $A$ is small:
\begin{align*}
	d(1,j)\gg1 \Rightarrow \epsilon_{j}\ll 1.
\end{align*}
In the case that members of the large group $A$ decide to collude, they can infer more information about the sensitive data $u$. Specifically, if a \textit{large} group $A$ averages the received responses $\{ y_{j}: j\in A \}$, the exact value of sensitive data $u$ is recovered. Indeed, composition theorem implies that only $\left( \sum_{j\in A} \epsilon_{j} \right)$-privacy of sensitive data $u$ is guaranteed. For a large group $A$, this privacy level becomes very loose.

\begin{figure} \begin{center}
\includegraphics[width=.6\linewidth]{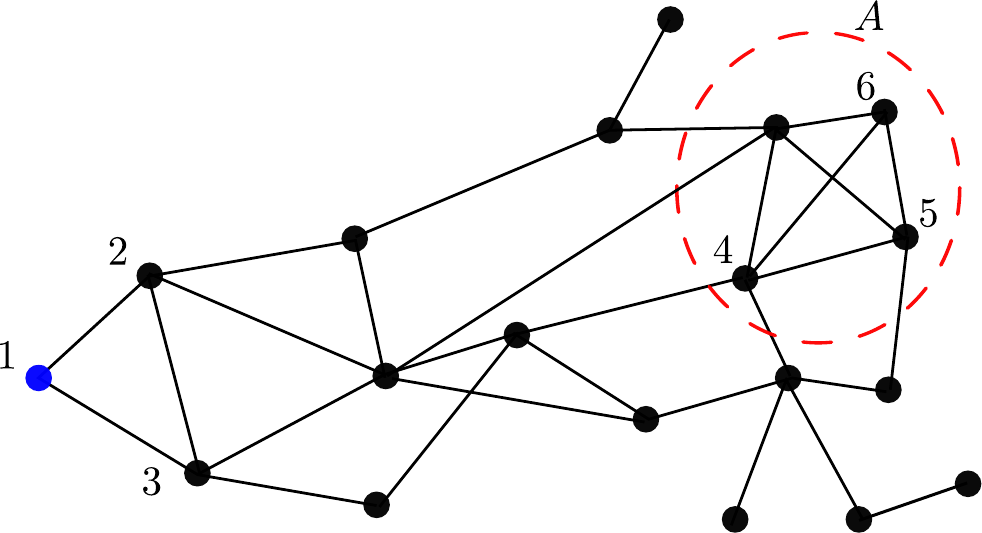}
\caption{User $1$ wants to share his sensitive data, such as his date of birth, in the a social network. Although, user $1$ has no privacy concerns when sharing this information with his close friends $2$ and $3$, he has gradually increasing privacy issues for other members of the network. Specifically, a group $A$ of distant users should not be able to collude and extract more information than what it is intended.} \label{fig:socialNetwork}
\end{center} \end{figure}

Our approach mitigates this issue. We assume that noisy versions of the private data are correlated and we design a mechanism that retains strong privacy guarantees. For real-valued sensitive data $u$, user $1$ samples $\{ v_{\epsilon} : \epsilon>0 \}$ from the stochastic process $\{ V_{\epsilon} : \epsilon>0 \}$, and responds to user $j$ with $y_{j} = u + v_{\epsilon_{j}}$, as shown in Figure \ref{fig:personalizedPrivacy}. In the case that a large group $A$ of users colludes, they are unable to extract much more information. Specifically, such a collusion renders individual's sensitive information at most $\left( \max_{j\in A} \epsilon_{j} \right)$-differential private. This privacy budget is significantly tighter than the one derived in the naive application of differential privacy and corresponds to the best information that a member of the group $A$ has. After all, if a close friend leaks sensitive information, it is impossible to revoke it.

\begin{figure} \begin{center}
\includegraphics[width=\linewidth]{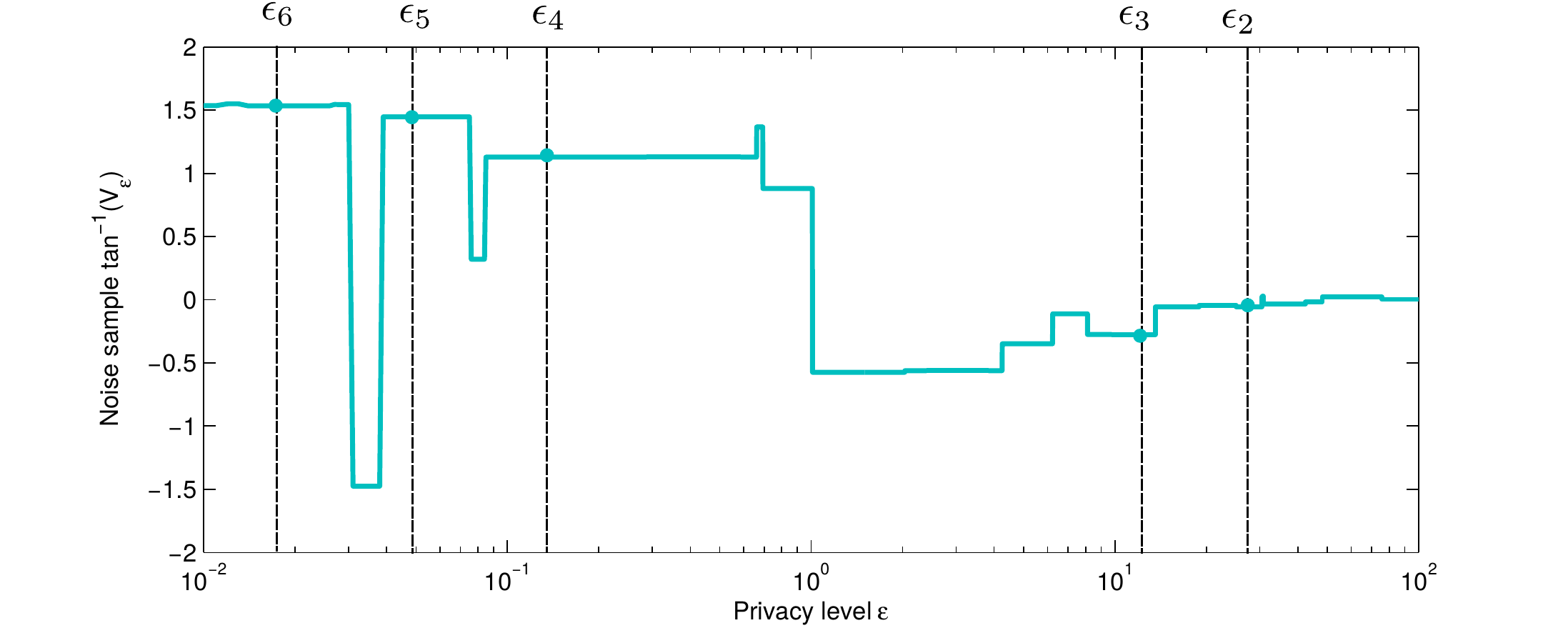}
\caption{User $1$ draws a single sample from the stochastic process $\{V_{\epsilon}\}_{\epsilon>0}$ and responds to user $i$ with $y_{i} = u + V_{\epsilon_{i}}$, where $\epsilon_{i}$ is the privacy level against user $i$. Eventually, having access to more responses $\{y_{i}\}_{i\in A}$ does not reveal more information about private data $u$ than the best response $\max_{\in A} \epsilon_{i}$.} \label{fig:personalizedPrivacy}
\end{center} \end{figure}
}

\section{Open Problems}
Finally, we conjecture that gradually releasing private data can be extended to any query and is, therefore, an intrinsic property of differential privacy. This conjecture is a key ingredient for the existence of a frictionless market of private data. In such a market, owners of private data can gradually agree to a rational choice of privacy level. Moreover, buying the exact private data is expected to be extremely costly. Instead, people may choose to buy private data in ``chunks'', in the sense of increasing privacy budgets.
We conjecture that gradually releasing sensitive data without loss in accuracy is feasible for a broader family of privacy-preserving mechanisms beyond mechanisms that approximate identity queries. This work was focused mechanisms which are defined on real space or sensitive data $\mathcal{U}=\reals^{n}$ under an $\ell_{1}$-norm adjacency relation, and approximate the identity query.

\longonly{
\section*{Acknowledgement}
The authors would like to thank Aaron Roth for providing useful feedback and suggesting the application of our results to Google's RAPPOR project.
}

\bibliographystyle{unsrt}
\bibliography{gradualPrivacy}

\end{document}